\documentclass[twoside,twocolumn,9pt]{article}
\usepackage{extsizes}
\usepackage[super,sort&compress,comma]{natbib} 
\usepackage[version=3]{mhchem}
\usepackage[left=1.5cm, right=1.5cm, top=1.785cm, bottom=2.0cm]{geometry}
\usepackage{balance}
\usepackage{mathptmx}
\usepackage{sectsty}
\usepackage{graphicx} 
\usepackage{lastpage}
\usepackage[format=plain,justification=justified,singlelinecheck=false,font={stretch=1.125,small,sf},labelfont=bf,labelsep=space]{caption}
\usepackage{float}
\usepackage{fancyhdr}
\usepackage{fnpos}
\usepackage[english]{babel}
\addto{\captionsenglish}{%
  \renewcommand{\refname}{Notes and references}
}
\usepackage{array}
\usepackage{droidsans}
\usepackage{charter}
\usepackage[T1]{fontenc}
\usepackage[usenames,dvipsnames]{xcolor}
\usepackage{setspace}
\usepackage[compact]{titlesec}
\usepackage{hyperref}
\usepackage{siunitx}
\definecolor{cream}{RGB}{222,217,201}

\begin{document}

\pagestyle{fancy}
\thispagestyle{plain}
\fancypagestyle{plain}{
%%%HEADER%%%
\renewcommand{\headrulewidth}{0pt}
}
%%%END OF HEADER%%%

%%%PAGE SETUP - Please do not change any commands within this section%%%
\makeFNbottom
\makeatletter
\renewcommand\LARGE{\@setfontsize\LARGE{15pt}{17}}
\renewcommand\Large{\@setfontsize\Large{12pt}{14}}
\renewcommand\large{\@setfontsize\large{10pt}{12}}
\renewcommand\footnotesize{\@setfontsize\footnotesize{7pt}{10}}
\makeatother

\renewcommand{\thefootnote}{\fnsymbol{footnote}}
\renewcommand\footnoterule{\vspace*{1pt}% 
\color{cream}\hrule width 3.5in height 0.4pt \color{black}\vspace*{5pt}} 
\setcounter{secnumdepth}{5}

\makeatletter 
\renewcommand\@biblabel[1]{#1}            
\renewcommand\@makefntext[1]% 
{\noindent\makebox[0pt][r]{\@thefnmark\,}#1}
\makeatother 
\renewcommand{\figurename}{\small{Fig.}~}
\sectionfont{\sffamily\Large}
\subsectionfont{\normalsize}
\subsubsectionfont{\bf}
\setstretch{1.125} %In particular, please do not alter this line.
\setlength{\skip\footins}{0.8cm}
\setlength{\footnotesep}{0.25cm}
\setlength{\jot}{10pt}
\titlespacing*{\section}{0pt}{4pt}{4pt}
\titlespacing*{\subsection}{0pt}{15pt}{1pt}
%%%END OF PAGE SETUP%%%

%%%FOOTER%%%
\fancyfoot{}
\fancyfoot[LO,RE]{\vspace{-7.1pt}\includegraphics[height=9pt]{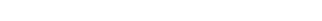}}
\fancyfoot[CO]{\vspace{-7.1pt}\hspace{13.2cm}\includegraphics{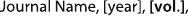}}
\fancyfoot[CE]{\vspace{-7.2pt}\hspace{-14.2cm}\includegraphics{head_foot/RF}}
\fancyfoot[RO]{\footnotesize{\sffamily{1--\pageref{LastPage} ~\textbar  \hspace{2pt}\thepage}}}
\fancyfoot[LE]{\footnotesize{\sffamily{\thepage~\textbar\hspace{3.45cm} 1--\pageref{LastPage}}}}
\fancyhead{}
\renewcommand{\headrulewidth}{0pt} 
\renewcommand{\footrulewidth}{0pt}
\setlength{\arrayrulewidth}{1pt}
\setlength{\columnsep}{6.5mm}
\setlength\bibsep{1pt}
%%%END OF FOOTER%%%

%%%FIGURE SETUP - please do not change any commands within this section%%%
\makeatletter 
\newlength{\figrulesep} 
\setlength{\figrulesep}{0.5\textfloatsep} 

\newcommand{\topfigrule}{\vspace*{-1pt}% 
\noindent{\color{cream}\rule[-\figrulesep]{\columnwidth}{1.5pt}} }

\newcommand{\botfigrule}{\vspace*{-2pt}% 
\noindent{\color{cream}\rule[\figrulesep]{\columnwidth}{1.5pt}} }

\newcommand{\dblfigrule}{\vspace*{-1pt}% 
\noindent{\color{cream}\rule[-\figrulesep]{\textwidth}{1.5pt}} }

\makeatother
%%%END OF FIGURE SETUP%%%

%%%TITLE, AUTHORS AND ABSTRACT%%%
\twocolumn[
  \begin{@twocolumnfalse}
{\includegraphics[height=30pt]{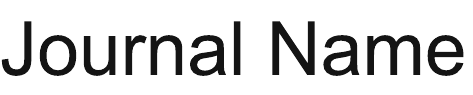}\hfill\raisebox{0pt}[0pt][0pt]{\includegraphics[height=55pt]{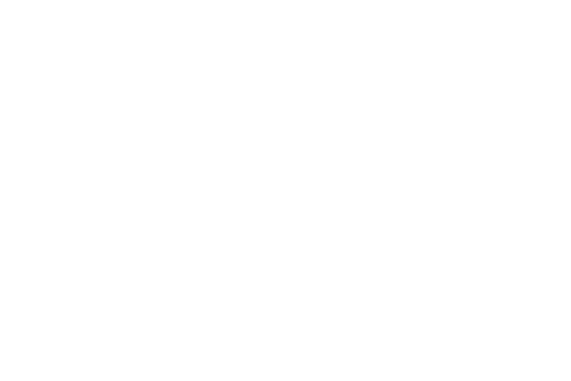}}\\[1ex]
\includegraphics[width=18.5cm]{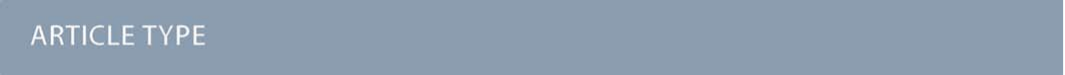}}\par
\vspace{1em}
\sffamily
\begin{tabular}{m{4.5cm} p{13.5cm} }

\includegraphics{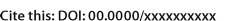} & \noindent\LARGE{\textbf{A \mbox{Multi-scale} Investigation of Aqueous Foams Stabilised by PNIPAM Microgels}} \\%Article title goes here instead of the text "This is the title"
\vspace{0.3cm} & \vspace{0.3cm} \\

 & \noindent\large{Joanne Zimmer,\textit{$^{a}$} Luca Mirau,\textit{$^{a}$} Kevin Gräff,\textit{$^{a}$} Ga\"{e}tan Barth,\textit{$^{a,b}$} Carina Schneider, \textit{$^{a}$} Vera A. N. A. Hesse, \textit{$^{a}$} Hayden Robertson \textit{$^{a}$} and Regine von Klitzing \textit{$^{a}$}$ ^{\ast}$} \\%Author names go here instead of "Full name", etc.

\includegraphics{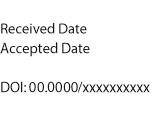} & \noindent\normalsize{Aqueous foams possess multiple structural motifs across different length scales: macroscopic foam, bubbles, foam films and the air/water interface.
In this study, macroscopic foams are generated by sparging gas through an aqueous dispersion of PNIPAM microgels which act as foam stabilisers due to their surface activity.
The stiffness of the microgels and thus their interfacial activity are tuned by variation of the \mbox{cross-linker} density. 
The effect of the \mbox{cross-linker} density and the microgel concentration on the resulting foam formation properties (foamability) and the foam stability are investigated.
A lower cross-linker density and a higher microgel concentration enhance the foamability, generate foams with smaller bubbles and higher liquid fractions, and increase the foam stability.
These observations are correlated with the microgel behavior at the single \mbox{air/water} interface examined by pendant drop tensiometry and Langmuir compression experiments as well as the mobility in single \mbox{free-standing} foam films determined using a Thin Film Pressure Balance. 
Our findings highlight good agreement across all length scales: increased foamability correlates with a faster decrease in surface tension, and higher foam stability with a higher surface elastic modulus of a microgel-covered single air/water interface and decreasing mobility in foam films.
} \\

\end{tabular}

 \end{@twocolumnfalse} \vspace{0.6cm}

  ]
%%%END OF TITLE, AUTHORS AND ABSTRACT%%%

%%%FONT SETUP - please do not change any commands within this section
\renewcommand*\rmdefault{bch}\normalfont\upshape
\rmfamily
\section*{}
\vspace{-1cm}

%%%FOOTNOTES%%%

\footnotetext{\textit{$^{a}$~Institut f\"ur Physik  Kondensierter Materie, Technische Universit\"at Darmstadt, D-64289 Darmstadt, Germany. E-mail: klitzing@smi.tu-darmstadt.de.}}
\footnotetext{\textit{$^{b}$~Univ. Bordeaux, CNRS, Bordeaux INP, ISM, UMR 5255, F-33400 Talence, France }}
\footnotetext{\dag~Electronic supplementary information (ESI) available: Height profiles of the microgels (MGs) determined by AFM with and without indentation; visualisation of the image skeletonisation process used to determine the bubble size; representative photographs captured by the CCD1 camera of the foam analyser device during foam formation and foam collapse; magnified view of the time evolution of the liquid fraction during foam formation of selected samples; an example illustrating the influence of the gas flow rate on the increase in foam volume during foam formation; and all fit parameters describing of the decrease in liquid fraction during the foam collapse. See DOI: }

%Please use \dag to cite the ESI in the main text of the article.
%If you article does not have ESI please remove the the \dag symbol from the title and the footnotetext below.
%\footnotetext{\dag~Supplementary Information available: [details of any supplementary information available should be included here]. See DOI: 00.0000/00000000.}
%additional addresses can be cited as above using the lower-case letters, c, d, e... If all authors are from the same address, no letter is required

%\footnotetext{\dag~Additional footnotes to the title and authors can be included \textit{e.g.}\ `Present address:' or `These authors contributed equally to this work' as above using the symbols: \ddag, \textsection, and \P. Please place the appropriate symbol next to the author's name and include a \texttt{\textbackslash footnotetext} entry in the the correct place in the list.}

%%%END OF FOOTNOTES%%%

%%%MAIN TEXT%%%%
%The main text of the article\cite{Mena2000} should appear here.

%\subsection{This is the subsection heading style}
%Section headings can be typeset with and without numbers.\cite{Abernethy2003}

%\subsubsection{This is the subsubsection style.~~} These headings should end in a full point.  

%\paragraph{This is the next level heading.~~} For this level please use \texttt{\textbackslash paragraph}. These headings should also end in a full point.

\section{Introduction}
Aqueous foams, dispersions of gas bubbles in a continuous liquid phase, are abundant in everyday life occurring in food products, cleaning and detergent agents as well as in personal care items \cite{Bamforth2023-bw, Schad2022-dp, Dari2025-qk, Mainkar2000-oa}.
In addition, foams are applied in firefighting and in large industrial applications such as froth or foam flotation \cite{Rakowska2018-bg, Matis1991-nh}.

Foams integrate multiple structural motifs across different length scales.
The smallest building blocks are foam films that separate two adjacent gas bubbles, or even the \mbox{air/water} interface.
Other motifs include Plateau borders, where three foam films meet, and nodes, connecting Plateau borders \cite{Koehler2000-rf, Hutzler1997-zm}.
Foam generation with water as the continuous phase requires a decrease in interfacial energy (surface tension) by the addition of a \mbox{surface-active} agent \cite{Pugh2016-bp}, which can be surfactants \cite{Boos2012-bw, Stubenrauch2009-wk, Arnould2018-zn, Lamolinairie2022-ew}, solid particles \cite{Binks2002-je, Horozov2008-so}, proteins \cite{Graff2022-wi, Graff2026-aj, Li2025-vz, Murray2004-iy, Engelhardt2013-jn}, polymers \cite{Keal2016-bg, D2SM01021F} or \mbox{polymer/surfactant mixtures} \cite{Bureiko2015-ky, Guzman2016-jr}.
Since foam generation is driven by a reduction in interfacial energy, the foamability, which describes the foam forming capacity of a dispersion, is often correlated with the adsorption kinetics of \mbox{surface-active} agents at freshly generated \mbox{air/water} interfaces \cite{Prins1992,Pugh1996, Engels2008-wr}.

Still, due to the large internal surface and the associated high surface energy, foams are thermodynamically unstable systems \cite{Drenckhan2015-jo}.
The main foam destabilisation mechanisms are gravitational drainage \cite{Kruglyakov2008-pi}, Ostwald ripening \cite{Briceno-Ahumada2017-xj} and the coalescence of adjacent bubbles \cite{Rio2014-yb}.
Drainage describes the \mbox{gravity-} and \mbox{capillary pressure-driven} liquid flow through the Plateau borders and nodes, and is strongly dependent on the liquid viscosity \cite{Koehler2000-rf, Saint-Jalmes2006-yu}.
It can be slowed down by increasing the viscous damping of the \mbox{surfactant-decorated} interface and by clogging of the nodes \cite{Pugh2016-bp,Varade2011-pn}.
Ostwald ripening, driven by the difference in Laplace pressure of adjacent bubbles, is decelerated by decreasing the gas solubility in water, and by reducing the gas permeability between the bubbles through the interfaces, \textit{e.g.} by increasing the foam film thickness or surface elastic modulus \cite{Wilson1996,Saint-Jalmes2005-tf, Saint-Jalmes2006-yu,CervantesMartinez2008-lq}.
Furthermore, increasing the stability of the foam films can slow down coalescence, as through film rupture the foam ultimately collapses \cite{Pugh2016-bp}.

Understanding foam stability is not only of fundamental interest but is also important for applications requiring a long foam lifetime.
In this respect, \mbox{particle-stabilised} systems, commonly referred to as Pickering systems, have become well established \cite{Han2023-tk,Zhao2022-hx, Zhang2022-ti}. 
In contrast to surfactants, particles adsorb at the interface in a practically irreversible manner with desorption energies up to several thousand $k_B T$, compared to only a few $k_B T$ for surfactants, enhancing the system stability \cite{Binks2002-je, Pugh2016-bp}.
This is a property also characteristic of microgels.
Microgels are \mbox{three-dimensional} polymeric networks of micrometer size, which due to their amphipilic nature, spontaneously and irreversibly adsorb at the \mbox{air/water} \cite{Zhang1999-lf} and \mbox{oil/water }interfaces \cite{Camerin2019-ys}.
However, in contrast to rigid colloidal particles, microgels possess the ability to swell and are deformable at liquid interfaces.
For this reason, they are especially interesting as they combine properties of both polymers and particles \cite{Plamper2017-kf}, giving rise to a more complex phase behavior.
Microgels made of poly(\textit{N}-isopropylacrylamide) (PNIPAM) have attracted particular attention as they exhibit a reversible volume phase transition when heated above approximately \SI{32}{\celsius}, referred to as the volume phase transition temperature (VPTT) \cite{Pelton2000-gw}.
The responsive nature of PNIPAM microgels can be transferred to the systems they stabilise, yielding, for example, \mbox{temperature-responsive} foams that can be destabilised on demand by raising the temperature above the VPTT \cite{Horiguchi2018-op, D2SM01021F}.
Furthermore, PNIPAM microgels are widely used model systems for the investigation of polymer adsorption at interfaces \cite{Fernandez-Rodriguez2021-av, Razavi2026-md, Kuk2024-gp, Vialetto2024-dq}.
When synthesised using the \mbox{one-batch} precipitation polymerisation method \cite{Pelton1986-gp}, the PNIPAM microgels yield a heterogeneous structure, \textit{i.e.} a densely \mbox{cross-linked} core and a loosely \mbox{cross-linked} shell with dangling polymer chains \cite{Stieger2004-jr, Boon2017-xo}.
In contrast to rigid particles, MGs can flatten and spread when adsorbed at the interface, resulting in a surface radius larger than in bulk \cite{Rey2020-ih}.
The degree of flattening is governed by the MG deformability which decreases with increasing \mbox{cross-liking} density and the surface tension of the liquid \cite{Burmistrova2011-im, Burmistrova2011-ud, Backes2017-id, Vialetto2022-qe}.

The collective microgel behavior as a function of \mbox{cross-linker} density on a model \mbox{air/water} interface such as on a Langmuir trough has been thoroughly investigated \cite{Pinaud2014,Rey2017,Tatry2023-og, Robertson2025-rk}.
Furthermore, the stabilising effect of microgels in single \mbox{free-standing} foam films has been subject of study \cite{Keal2016-bg, Cohin2013, D2SM01021F}.  
A systematic study investigating the influence of microgel properties such as \mbox{cross-linker} density and microgel concentration on the macroscopic foam characteristics has not yet been conducted.
Therefore, we herein present a \mbox{multi-scale} approach towards understanding the relationship between different \mbox{microgel-stabilised} \mbox{air-water} interfaces and the formation and stability of macroscopic foams. 
Here, the macroscopic foams are generated by sparging gas through microgel dispersions and analysed using a commercial foam analyser device, also used for the investigation of foamability and stability of \mbox{surfactant-stabilised} foams \cite{Boos2012-bw, Boos2013-us, Stubenrauch2009-wk, Preisig2019-mm} also in combination with superchaotropic \mbox{nano-ions} \cite{ Hohenschutz2021-hb}, {protein-stabilised} foams \cite{Schmidt2010-lb, Rullier2010-wb} and {particle-stabilised} foams \cite{Choi2020-ql} . 
Both the \mbox{cross-linker} density in the microgels and the microgel concentration in the foaming dispersion are varied.
The experiments are complimented by investigation of \mbox{microgel-stabilised} single \mbox{air/water} interfaces such as those on a Langmuir trough and in pendant drop tensiometry, as well as single \mbox{free-standing} foam films in which two \mbox{air/water} interfaces are located in close proximity to each other.

\section{Experimental}
\label{Exp}
\subsection{Chemicals}
\textit{N}-isopropylacrylamide (NIPAM, $(\geq99\%)$), \textit{N,N}'-methylenbisacrylamide (BIS, $99\%$) and sodium chloride (NaCl, $99.99\%$) were purchased from Sigma-Aldrich (Merck, Darmstadt, Germany) and used as received. 
Ethanol (EtOH, $(\geq99.9\%)$) was purchased from Carl Roth GmbH + Co. KG (Karlsruhe, Germany).
2,2'-azobis(2-methyl-propanimidamide) dihydrochloride (AAPH, $98\%$) was purchased from Cayman Chemical Company (Cayman Chemical, USA). Purified water (specific resistivity $\SI{18.2}{\mega\Omega\cdot\centi\meter}$ at \SI{25}{\celsius}) from a Milli-Q purification system (Merck KGaA, Darmstadt, Germany) was used. 

\subsection{Microgel synthesis}
The PNIPAM microgels (MGs) were synthesised by \mbox{surfactant-free} precipitation polymerization \cite{Pelton1986-gp}.
Here, NIPAM and BIS were dissolved in \SI{120}{mL} of \ce{H2O} and transferred into a home-built, double-walled glass reactor. 
In line with previous works\cite{D2SM01021F, Stock2024-zy, Robertson2025-rk}, MGs with the targeted \mbox{cross-linker} densities of \SI{2}{mol\%}, \SI{5}{mol\%} and \SI{10}{mol\%} BIS were obtained by variation of the NIPAM/BIS ratio respectively while the sum of NIPAM and BIS remained constant at \SI{20}{mmol} (see Tab.~\ref{tbl:MG_synthesis}).
Next, the mixture was degassed with \ce{N2} at \SI{80}{\celsius} for \SI{60}{min} under constant stirring at 500 rpm. 
For initiation of the polymerisation reaction \SI{0.125}{mmol} of AAPH dissolved in \SI{1}{\milli\liter} \ce{H2O} were injected into the reaction mixture under constant stirring at \SI{1000}{rpm}. 
Stirring was terminated after \SI{60}{min} and the turbid solution was rapidly cooled down in an ice bath. 
For removal of undesired synthesis residues, the MG suspension was dialysed for 10 days against Milli-Q \ce{H2O}, with a daily water change (against \SI{50}{\liter} in total). 
Following this, the MG suspension was subjected to centrifugation (10350 rpm). 
This was performed 5 times for \SI{2}{h} for the low \mbox{cross-linked} MGs \textit{i.e.} \SI{2}{mol\%} BIS and 3 times for \SI{30}{\min} for the higher \mbox{cross-linked} MGs, \textit{i.e.} \SI{5}{mol\%} BIS and \SI{10}{mol\% } BIS. 
After each centrifugation step the supernatant was carefully removed and replaced by an equal amount of \ce{H2O} before the MGs were fully redispersed.
Finally, the obtained MG suspension was \mbox{freeze-dried} and the dried MG were stored at \SI{-20}{\celsius} before usage.
The synthesised MGs and their chemical composition are listed in Tab.~\ref{tbl:MG_synthesis}.
The sample notation indicates the \mbox{cross-linker} density in \si{\mol\percent}, given by the number following "MG".
For example, MG2 denotes MGs prepared with \SI{2}{mol\%} BIS.
\begin{table}[h]
\small
  \caption{NIPAM and BIS concentrations used for synthesis of microgels with varying \mbox{cross-linker} (BIS) density}
  \label{tbl:MG_synthesis}
  \begin{tabular*}{0.48\textwidth}{@{\extracolsep{\fill}}llll}
    \hline
    Microgel & $c(\mathrm{BIS})$ / \si{\mol\percent} & $n(\mathrm{NIPAM})$ / \si{\milli\mol} & $n(\mathrm{BIS})$ / \si{\milli\mol}\\
    \hline
    MG2 & 2 & 19.6 & 0.4 \\
    MG5 & 5 & 19.0 & 1.0 \\
    MG10 & 10 & 18.0 & 2.0 \\
    \hline
  \end{tabular*}
\end{table}

\subsection{Microgel characterisation}
\subsubsection{Dynamic light scattering and \mbox{zeta potential}.}
The MGs hydrodynamic radius ($R_H$) was determined by dynamic light scattering performed on the \mbox{multi-angle} device from LS Instruments (LS Instruments AG, Fribourg, Switzerland).
The zeta potential ($\zeta$) was measured with a Zetasizer Nano (Malvern Panalytical, UK) under the Smoluchowski approximation.
For both measurements the sample concentration was \SI{0.005}{wt\percent} in \ce{H2O}.
The temperature was varied between \SI{20}{\celsius} and \SI{50}{\celsius} to obtain $R_H$ respective $\zeta$ as a function of temperature.
The swelling ratio ($SR$) was calculated as the volume ratio of the MGs at \SI{20}{\celsius} and \SI{50}{\celsius} according to Eq.~\ref{eq:SR},

\begin{equation}
\label{eq:SR}
  SR = \frac{V(\SI{20}{\celsius})}{V(\SI{50}{\celsius})}
  = \frac{R_\mathrm{H}(\SI{20}{\celsius})^3}{R_\mathrm{H}(\SI{50}{\celsius})^3}
\end{equation}

\noindent where $V$ is the volume and $R_\mathrm{H}$ the hydrodynamic radius of the MGs at \SI{20}{\celsius} and \SI{50}{\celsius} respectively.

\subsubsection{Compression isotherms and Langmuir--Blodgett deposition.}

The compression isotherms of the MG monolayer were recorded using an RK1 standard Langmuir trough ($A_{\text{min}} = \SI{13.18}{\square\centi\meter}, \quad A_{\text{max}} = \SI{207.47}{\square\centi\meter}$, Riegler \& Kirstein GmbH, Potsdam, Germany). 
A specific volume of the MG sample (\SI{1}{wt\%} in \ce{H2O}, \SI{20}{vol\%} EtOH) was spread drop wise on the air/water interface and was left to equilibrate for \SI{30}{\min} prior to compression to allow for evaporation of EtOH and to ensure an equilibrated MG film at the air/water interface.
The film was then compressed with a constant barrier speed of \SI{12.15}{\square\milli\meter\per\second}.
The surface pressure ($\varPi_\mathrm{s}$) was determined by a Wilhelmy tensiometer equipped with blotting paper.
In the resulting  $\varPi_\mathrm{s}(A)$ isotherms, with $A$ being the area of the air/water interface, $A$ was normalised to the MG mass spread.
To account for artifacts within the isotherms, interpolating cubic splines were then employed to describe  the $\varPi_\mathrm{s}(A)$ isotherms.
Subsequently, the surface elastic modulus ($E_\mathrm{s}$) was calculated by Eq.~\ref{eq:E_s} according to the work of Tatry \textit{et al.} \cite{Tatry2023-og}.

\begin{equation}
\label{eq:E_s}
  E_\mathrm{s} = -\frac{\mathrm{d}\varPi_\mathrm{s}}{\mathrm{d}\ln A}
\end{equation}

\mbox{Langmuir--Blodgett} deposition of the MGs at the air/water interface was carried out at a defined surface pressure of $\varPi_\mathrm{s} = \SI{0.5}{\milli\newton\per\meter}$ on a silicon substrate with a natural oxide layer.
The procedure is described in detail elsewhere \cite{Robertson2025-rk}.
In short, the MG monolayer was prepared in the same way as explained above and compressed until $\varPi_\mathrm{s} = \SI{0.5}{\milli\newton\per\meter}$.
Then the system was left to equilibrate for another \SI{30}{\min}, after which the silicon wafer, which had been installed beforehand, was lifted through the interface with a constant speed of \SI{0.1}{\milli\meter\per\minute} at a \SI{45}{\degree} angle.
A constant lateral pressure was ensured by a continuous adjustment of the barrier positions through the trough's feedback system.
The slightly inclined wafers dried during the deposition process due to the slow lifting movement.
All experiments were carried out at a constant temperature of \SI{20}{\celsius}.

\subsubsection{Atomic force microscopy.}
The apparent elastic modulus ($E_\mathrm{MG}$) of the MGs in \ce{H2O} was determined by atomic force microscopy (AFM) indentation measurements of the MG samples prepared by Langmuir--Blodgett deposition. 
These measurements were performed on the Nanowizard 4 XP AFM (JPK, Berlin, Germany) using BL-AC40TS cantilevers with a nominal spring constant of $k = \SI{0.09}{\newton\m^{-1}}$ and an average resonance frequency of $f_{\text{res}} = \SI{110}{\kilo\hertz}$ (Oxford Instruments, UK).
Before each measurement the exact spring constant of the cantilever was identified by the \mbox{contact-based} calibration method on a hard silicon wafer.
The MG sample, as well as the cantilever, were submerged into \ce{H2O} and left to equilibrate for at least \SI{15}{\min} at \SI{20}{\celsius} before starting the measurement. 
By choosing the QI Advanced Imaging mode of the AFM device, a \mbox{force-distance} curve was recorded at each pixel of the image, resulting in the generation of a force map.
The maximum indentation force was set to \SI{3}{\nano\newton}.
$E_\mathrm{MG}$ was determined by fitting the curves with the Hertz model \cite{Hertz1882} (Eq.~\ref{eq:Hertz_model}),
\begin{equation}
F = \frac{2}{\pi} \tan(\alpha) \frac{E_\mathrm{MG}}{(1 - \nu)^2} \delta^2
\label{eq:Hertz_model}
\end{equation}

\noindent with $F$ the force applied by the conical indenter with a \mbox{half-cone} angle $\alpha$ ($\alpha = \SI{20}{\degree}$), $E_\mathrm{MG}$ the elastic modulus, $\nu$ the Poisson ratio ($\nu = 0.4$ \cite{Voudouris2013-fe}) and $\delta$ the indentation depth. 
The fit region was set to \SI{20}{\percent} of the maximum indentation force (\SI{0.6}{\nano\newton}).
$E_\mathrm{MG}$ was determined as the average of seven MGs.
Additionally, the MG height ($h$(AFM,\SI{20}{\celsius})) was extracted from the contact point offset, defined as the vertical distance between the substrate and MG contact point prior to indentation.
The values of $h$(AFM,\SI{20}{\celsius}) at the center of the MGs are listed in Tab.~\ref{tbl:MG_properties}.
Furthermore, the indentation depth was determined by evaluating the MG height at \SI{0.6}{nN} of applied force.
Depending on the MGs elastic modulus $E_\mathrm{MG}$, an indentation force of \SI{0.6}{nN} corresponds to an indentation depth of $100\pm 14$ \si{\nano\meter} (MG2), $ 66\pm8$ \si{\nano\meter} (MG5) and $34\pm7$ \si{\nano\meter} (MG10) at the center of the MGs.
The full curves showing the MG height as a function of MG \mbox{cross-section}, both with and without indentation are presented in Fig.~S1 in the ESI\dag.

\subsubsection{Pendant drop tensiometer.}
The kinetics of the surface tension ($\gamma(t)$) was measured with a drop shape analyzer OCA 20 (DataPhysics Instruments, Filderstadt, Germany) using the pendant drop technique (drop volume $\SI{4.5}{\mu\liter}$).
To avoid evaporation the syringe tip was placed in a sealed glass cuvette with a water vapor saturated atmosphere.
The surface tension was obtained by fitting the experimentally determined pendant drop contour with profiles calculated from the \mbox{Young--Laplace} equation.
To reduce the influence of thermal fluctuations, surface tension values were averaged over three adjacent data points.
The MG samples were investigated at concentrations of \SI{0.01}{wt\%}, \SI{0.025}{wt\%} and \SI{0.05}{wt\%} in \ce{H2O}.
The experimental sample notation MG$b$\_$c$ indicates the examined MG \mbox{cross-linker} density ($b$) and concentration ($c$).
For example, MG2\_0.05 corresponds to an experiment with \SI{0.05}{wt\%} of MG2.

\subsection{Mobility in \mbox{free-standing} foam films}
Single \mbox{free-standing} foam films prepared from MG dispersions were investigated using a thin film pressure balance (TFPB) with the porous plate technique \cite{Mysels1966, Scheludko1960}.
The procedure is described in detail by Gräff \textit{et al.} \cite{Graff2022-wi, Graff2026-aj}.
In short, the film holder in our \mbox{custom-built} setup consists of a porous glass plate (porosity P16 (ISO 4793), pore size \SIrange{10}{16}{\micro\meter}) possessing a drilled hole (diameter \SI{1}{\milli\meter}) which is connected to a bent glass capillary tube.
The film holder is positioned in a sealed stainless steel chamber, also containing a reservoir of the aqueous MG sample investigated to ensure a saturated atmosphere and impede film drying.
For equilibration of the porous glass plate, the film holder was immersed into the MG sample (\SI{0.3}{wt\%} in \ce{H2O}) for at least \SI{2}{\hour}  prior to each measurement.
The film holder was then raised from the MG sample solution and equilibrated for another \SI{30}{\minute} in the saturated atmosphere in the pressure chamber.
During the measurement the pressure inside the pressure chamber is increased and a single foam film forms in the hole of the porous glass plate.
The disjoining pressure $\varPi_\mathrm{d}$ is calculated with Eq.~\ref{eq:disj_pressure},
\begin{equation}
\varPi_\mathrm{d} = P_\mathrm{g} - P_\mathrm{r} + \frac{2\gamma}{r} - \Delta \rho g h_\mathrm{c}
    \label{eq:disj_pressure}
\end{equation}
with $P_\mathrm{g}$ being the pressure applied to the pressure chamber, $P_\mathrm{r}$ the ambient reference pressure, $\gamma$ the surface tension of the sample solution, $r$ the radius of the glass capillary tube, $\Delta\rho$ the density difference between solution and air, $g$ the gravitational constant and $h_\mathrm{c}$ the height of the liquid inside the glass capillary tube above the foam film.
The foam films were investigated with an optical microscope in top view (Nikon) connected to a camera (JAI GO-2400C-USB, \SI{5.86}{\micro\meter} $\times$ \SI{5.86}{\micro\meter} pixel size, Stemmer Imaging, Germany ).
The resulting foam films were colorful (thickness greater than \SI{100}{\nano\meter}) and exhibited distinguishable features \textit{i.e.} regions of higher contrast \cite{D2SM01021F}.
To identify the mobility in the foam films three features at different positions in an image corresponding to an early stage of foam film formation ($\varPi_\mathrm{d},_0$) were selected, their coordinates determined and set as reference position ($x_0$, $y_0$).
The feature displacement ($\Delta r$) was obtained by tracking the position of the features throughout the foam films life span.
Feature positions were determined with a quadratic pixel array covering the whole feature.
The corresponding pixel coordinates ($x$, $y$) were identified with the ImageJ software.
With the edge length of one quadratic pixel corresponding to \SI{0.67}{\micro\meter}, $\Delta r$ can be calculated according to Eq.~\ref{eq:delta_r},

\begin{equation}
  \Delta r(\varPi_\mathrm{d}) = 0.67 \si{\micro\meter} \cdot
\sqrt{ \bigl(x(\varPi_\mathrm{d}) - x_0(\varPi_\mathrm{d},_0)\bigr)^2
     + \bigl(y(\varPi_\mathrm{d}) - y_0(\varPi_\mathrm{d},_0)\bigr)^2 } 
  \label{eq:delta_r}   
\end{equation}

\noindent with $x(\varPi_\mathrm{d})$ and $y(\varPi_\mathrm{d})$ being the feature positions at the corresponding disjoining pressure $\varPi_\mathrm{d}$, expressed as pixel coordinates.
Each experiment was repeated three times at a constant temperature of \SI{22}{\celsius}.

\subsection{Foaming experiments}
\begin{figure}[h]
\centering  \includegraphics[height=10cm]{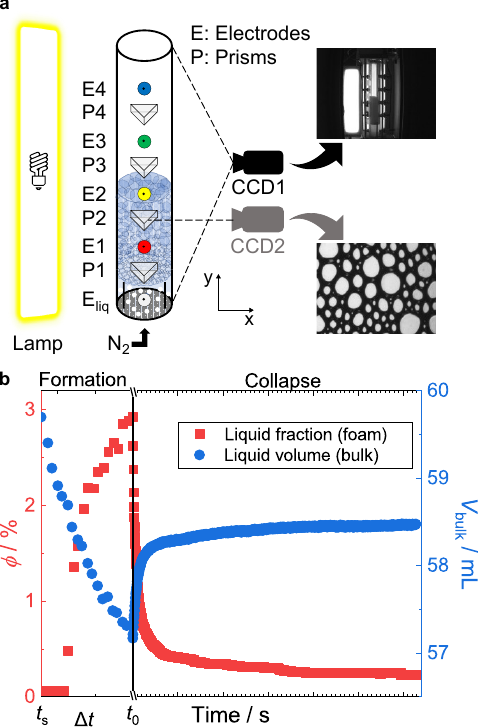}
  \caption{a) Schematic representation of the foam analyser device used for generation and investigation of the \mbox{MG-stabilised} foams. b) Time evolution of the liquid fraction in the foam ($\phi$) and the bulk liquid volume at the bottom of the tube ($V_\mathrm{bulk}$) in a typical foaming experiment. The characteristic points in time $t_s$, $t_0$ and the time period $\Delta t$, as well as the experiment sections associated with foam formation and collapse are indicated.}
  \label{fgr:foamscan}
\end{figure}

Foaming experiments were carried out using a commerically available foam analyser device (TECLIS Scientific, Civrieux d’Azergues, France) (cylinder with one flattened side, inner glass tube diameter of \SI{35}{\milli\meter}).
A schematic representation of the device is shown in Fig.~\ref{fgr:foamscan}a).
It consists of a glass tube decorated with sets of electrodes (E) located at different heights as well as four prisms (P).
The entire tube is well illuminated by a lamp.
One camera (CCD1) captures the whole foam analyser tube, whereas a second camera (CCD2) is focused on one of the prisms to give a close up image of the foam, showing the foam bubbles at the tube wall.

The foams were generated by sparging nitrogen (\ce{N2}) with a flow rate ($Q$) of $Q= \SI{100}{\milli\liter\min^{-1}}$ through a porous silica frit (porosity P16 (ISO 4793), pore size \SIrange{10}{16}{\micro\meter}) located at the bottom of the tube and \SI{60}{\milli\liter} of MG dispersion ($V_\mathrm{bulk}$).
Before each measurement, the system was calibrated according to the manufactures protocol by determination of the liquid conductance over liquid volume ($E$\textsubscript{liq}) by use of the vertical electrodes.

For investigation of the effect of \mbox{cross-linker} density, foams were generated from \SI{0.05}{wt\%} aqueous dispersions of MG2, MG5 and MG10.
To investigate the effect of MG concentration, foams were generated from \SI{0.01}{wt\%}, \SI{0.025}{wt\%} and \SI{0.05}{wt\%} aqueous dispersions of MG2.
As the MGs are only slightly charged, each sample additionally contained \SI{1}{mM} sodium chloride (NaCl) as an electrolyte to allow for adequate tracking of the electrical conductivity.
The experiment notation MG$b$\_$c$ contains information about the MG \mbox{cross-linker} density ($b$) and concentration ($c$).
For example the sample name MG2\_0.05 indicates an experiment with \SI{0.05}{wt\%} of MG2.

The liquid fraction ($\phi$) in the foam was obtained with \mbox{electrode 1} (\mbox{E1}, $h = \SI{90}{\milli\meter}$) by measuring the electric conductivity from which $\phi$ is calculated based on the empirical relation by Feitosa \textit{et al.} \cite{Feitosa2005-xu}.

Images of the foam bubbles were taken using the CCD2 camera at the height of \mbox{prism 2} (\mbox{P2}, $h = \SI{105}{\milli\meter}$).
From these images the Sauter diameter of the foam bubbles was determined by the Cell Size Analysis (CSA) software provided by TECLIS Scientific (Civrieux d’Azergues, France) and ImageJ.
First, the images were treated with ImageJ to obtain a \mbox{water-free} skeleton by reducing the dark surface Plateau borders of the prior binarised image to \mbox{single-pixel} width (see Fig.~S2 in the ESI\dag).
This is required because the Plateau borders at the tube surface have a \mbox{cross-section} about three times larger than those inside the foam, which otherwise would distort the proper estimation of the bubble size distribution \cite{Boos2013-us}.
Subsequently, the skeletonised images were evaluated with the CSA software.

The target foam volume ($V_\mathrm{foam}$) to be generated  was \SI{100}{\milli\liter} and was automatically detected by the foam analyser device.
After formation, the foams did not collapse uniformly with a \mbox{well-defined} \mbox{gas/foam} interface.
Furthermore, foam fragments occasionally remained attached to the tube walls (see Fig.~S3 and S4 in the ESI \dag).
As a result, it was not possible to reliably determine the foam volume during the foam collapse.
For this reason, the foam volume is displayed only during foam formation. 

To allow for comparison between different experiments the following times and time periods were defined in accordance with Boos \textit{et al.} \cite{Boos2013-us}:
$t_\mathrm{s}$ is the experiment starting point at which the gas flow is started.
At $t_\mathrm{0}$ the target foam volume of $V_\mathrm{foam}=\SI{100}{\milli\liter}$ is reached and the gas flow is stopped.
The period of time between $t_\mathrm{s}$ and $t_\mathrm{0}$ is called $\Delta t$ and is associated with foam formation.
Investigation of foam stability starts after $t_\mathrm{0}$. 
Therefore, $t_\mathrm{0}$ is always set to \SI{0}{\second}, allowing for the investigation of foam decomposition from a defined starting point.

For clarification, $t_\mathrm{s}$, $t_\mathrm{0}$ and $\Delta t$ in a typical foaming experiment are shown in Fig.~\ref{fgr:foamscan}b), depicting the time evolution of the liquid fraction ($\phi$) in the foam  and the bulk liquid volume at the bottom of the tube ($V_\mathrm{bulk}$) during foam formation and collapse.
During foam formation $\phi$ increases, while $V_\mathrm{bulk}$ decreases as liquid is transferred into the foam.
Foam formation is completed at $t_\mathrm{0}$.
During foam collapse, \textit{i.e.} after $t_\mathrm{0}$, $\phi$  decreases, while $V_\mathrm{bulk}$ increases as liquid is recovered.

All experiments were carried out at a constant temperature of \SI{20}{\celsius}, ensured by a tempered water jacket surrounding the foam analyser apparatus.
Each foam was generated at least three times from a fresh foaming solution to obtain error bars, with the glass tube being thoroughly rinsed with water between each measurement.
To enable calculation of \mbox{point-wise} mean values across repeated measurements, the $\phi$ data were linearly interpolated onto a common temporal grid with \SI{3}{\second} intervals and the $V_\mathrm{foam}$ data were linearly interpolated onto a common temporal grid with \SI{0.5}{\second} intervals.
Mean values and error bars were calculated only for time points at which data were available for all repetitions.

\section{Results}
\subsection{Microgel properties}
First, the MG properties are investigated as a function of \mbox{cross-linker} density.
The hydrodynamic radii $R_\mathrm{H}$ and zeta potentials $\zeta$ of MG2, MG5 and MG10 at \SI{20}{\celsius} and \SI{50}{\celsius}, the swelling ratio $SR$, and the \mbox{AFM-measured} heights at the center of the MGs in \ce{H2O} at \SI{20}{\celsius} $h$(AFM,\SI{20}{\celsius}), are listed in Tab.~\ref{tbl:MG_properties}.
For all MG systems, $R_\mathrm{H}$ decreases significantly from \SI{20}{\celsius} to \SI{50}{\celsius} owing to the \mbox{well-known} volume phase transition (VPT) of PNIPAM MGs \cite{Pelton2000-gw}.
The swelling ratio $SR$, derived from $R_\mathrm{H}$ at \SI{20}{\celsius} and \SI{50}{\celsius}, decreases with increasing \mbox{cross-linker} density, reflecting the higher degree of \mbox{cross-linking} in these MGs.
Furthermore, the $\zeta$ potential increases significantly from \SI{20}{\celsius} to \SI{50}{\celsius}, which can be attributed to the decrease in MG surface area during the VPT while the total charge in the individual MG system remains unchanged \cite{Daly2000}. 
The positive charge arises from the cationic initiator (AAPH) present in the MGs.
The MG height $h$(AFM,\SI{20}{\celsius}) increases with increasing \mbox{cross-linker} density, as MGs with lower \mbox{cross-linker} density are more deformable and, therefore, exhibit a larger spreading on the hydrophilic substrate \cite{Rey2017, D2SM01021F}. 
The complete height profiles as a function of MG \mbox{cross-section} without indentation (\SI{0}{\nano\newton}) are provided in Fig.~S1 in the ESI\dag.

\begin{table*}
\small
  \caption{Hydrodynamic radii $R_\mathrm{H}$ at \SI{20}{\celsius} and \SI{50}{\celsius}, the swelling ratio $SR$, $\zeta$ potentials at \SI{20}{\celsius} and \SI{50}{\celsius}, and microgel heights in \ce{H2O} at \SI{20}{\celsius} determined by AFM $h${(AFM,\SI{20}{\celsius}}) of MG2, MG5 and MG10}
  \label{tbl:MG_properties}
  \begin{tabular*}{\textwidth}{@{\extracolsep{\fill}}llllllll}
    \hline
    Microgel & $R_\mathrm{H}$ (\SI{20}{\celsius}) / \si{\nano\meter} & $R_\mathrm{H}$ (\SI{50}{\celsius}) / \si{\nano\meter} & $SR$ & $\zeta$ (\SI{20}{\celsius}) / \si{\milli\volt}  & $\zeta$ (\SI{50}{\celsius}) / \si{\milli\volt} & $h${(AFM,\SI{20}{\celsius}}) / \si{\nano\meter}\\
    \hline
    MG2 & $312 \pm 3$ & $143 \pm 1$  & $10.4 \pm 0.4 $ & $3 \pm$ 4 & $40 \pm 4$ & $220 \pm 11$\\
    MG5 & $273 \pm 2$ & $135 \pm 1$ & $8.3 \pm 0.3 $ & $7 \pm 4$ & $38 \pm 5$ & $257 \pm 6$ \\
    MG10 & $282 \pm 3$ & $153 \pm 3$ & $6.2 \pm 0.4 $ & $21 \pm 4$& $32 \pm 6$& $393 \pm 5$ \\
    \hline
  \end{tabular*}
\end{table*}

\subsection{Microgels at a planar air/water interface }
As a first step in the \mbox{multi-scale} investigation of \mbox{MG-stabilised} aqueous foams, the behaviour of the MGs at planar, single \mbox{air/water} interfaces was examined using Langmuir trough experiments and pendant drop tensiometry.
In this regard, the elastic modulus $E_\mathrm{MG}$ of the individual MGs and their morphology, both determined by AFM, are related to the resulting interfacial properties.

\begin{figure*}[htb]
 \centering
 \includegraphics[height=10cm]{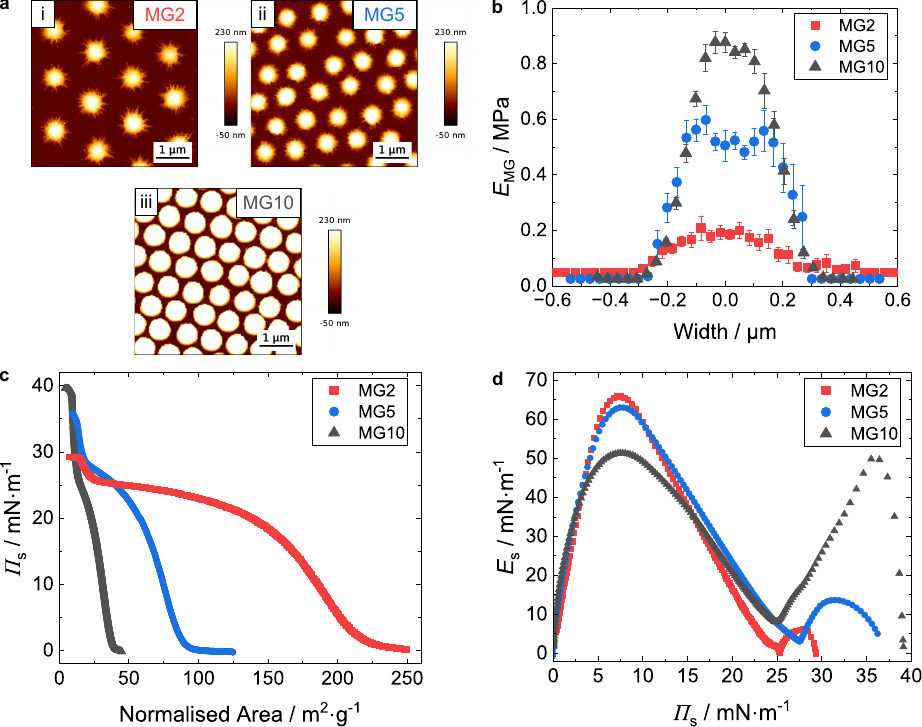}
 \caption{a) AFM micrographs of (i) MG2, (ii) MG5 and (iii) MG10 deposited on a silicon wafer using the Langmuir--Blodgett technique at a defined surface pressure of $\varPi_\mathrm{s} = \SI{0.5}{\milli\newton\per\meter}$. Scans in \ce{H2O}. b) Elastic modulus $E_\mathrm{MG}$ over MG cross-section for MG2, MG5 and MG10. c) Langmuir compression isotherms (surface pressure $\varPi_\mathrm{s}$ as a function of normalised area) of the MG monolayers of MG2, MG5 and MG10. d) Surface elastic modulus $E_\mathrm{s}$ as a function of surface pressure $\varPi_\mathrm{s}$ derived from Eq.~\ref{eq:E_s} for MG2, MG5 and MG10. All measurements were conducted at a temperature of  \SI{20}{\celsius}.}
 \label{fgr:AFM_MG}
\end{figure*}

Fig.~\ref{fgr:AFM_MG}a) shows AFM micrographs of MG2, MG5 and MG10 after sample preparation by \mbox{Langmuir--Blodgett deposition} at a surface pressure of $\varPi_\mathrm{s} = \SI{0.5}{\milli\newton\per\meter}$.
The tendency to form a hexagonal densely packed structure becomes obvious.
Furthermore, MGs with the lowest \mbox{cross-linker} density (MG2) exhibit a distinct \mbox{core-shell} structure, which becomes less pronounced with increasing \mbox{cross-linker} density. 
In addition, a higher number of MGs per unit area is required to obtain the surface pressure of $\varPi_\mathrm{s} = \SI{0.5}{\milli\newton\per\meter}$ with increasing \mbox{cross-linker} density.
Generally, MGs deform when adsorbed at liquid interfaces, with a larger diameter than in bulk \cite{Rey2020-ih}. 
However, the capability for deformation decreases with increasing \mbox{cross-linker} density, also reflected in the lower swelling ratios shown in Tab.~\ref{tbl:MG_properties}, which requires more MGs to cover a given surface area.

The $E_\mathrm{MG}$ as a function of MG \mbox{cross-section} of MG2, MG5 and MG10 is displayed in Fig.~\ref{fgr:AFM_MG}b).
With increasing \mbox{cross-linker} density the $E_\mathrm{MG}$ increases \textit{i.e.} the MGs become stiffer, which is also known from other experimental studies in the literature and a result of the higher degree of \mbox{cross-linking} \cite{Burmistrova2011-im, Backes2018, D2SM01021F}. 
The bell shape of the $E_\mathrm{MG}$ over the \mbox{cross-section} is attributed to the MGs \mbox{core-shell} structure with a higher \mbox{cross-linked}, harder core and a more loosely \mbox{cross-linked}, softer shell, since BIS has a higher reaction rate than NIPAM, the core has a higher BIS content than the shell \cite{Wu1994-ci}.
The height profiles as a function of MG \mbox{cross-section} when indented with a force of \SI{0.6}{\nano\newton}, the maximum indentation force for the fit of the Hertz model, are shown in Fig.~S1 in the ESI\dag.
The absolute indentation depth decreases with increasing \mbox{cross-linker} density because the MGs become stiffer.

Fig.~\ref{fgr:AFM_MG}c) depicts the Langmuir compression ($\varPi_\mathrm{s}(A)$) isotherms of MG2, MG5, MG10.
Consistent with previous literature \cite{Picard2017, Rey2017, Tatry2023-og}, the surface pressure $\varPi_\mathrm{s}$ begins to increase at larger surface areas for MGs with lower \mbox{cross-linker} density.
Again, this is linked to the stronger deformation of the lower \mbox{cross-linked} MGs adsorbed at the \mbox{air/water} interface, thus, covering a larger surface area. 
The surface elastic modulus $E_\mathrm{s}$ as a function of $\varPi_\mathrm{s}$ calculated from the Langmuir isotherms by means of Eq.~\ref{eq:E_s} are shown in Fig.~\ref{fgr:AFM_MG}d).
The curves display the classical shape reported in literature \cite{Pinaud2014, Picard2017, Tatry2023-og}.
The first maximum stems from the interacting MG shells; its height increases with lower \mbox{cross-linker} density due to a larger density of stretched dangling chains \cite{Pinaud2014}.
The second maximum, originating from \mbox{core-core} interactions, becomes more pronounced with increasing \mbox{cross-linker} density, consistent with the higher $E_\mathrm{MG}$ of the individual MGs \cite{Tatry2023-og}.
Forced compression \textit{e.g.} by means of a Langmuir trough enables access to the full range of surface coverages, whereas in spontaneous adsorption MGs adsorb below the second elastic maximum \cite{Pinaud2014, Tatry2019-hd, Tatry2023-og}.

\begin{figure}[htb]
\centering
  \includegraphics[height=10cm]{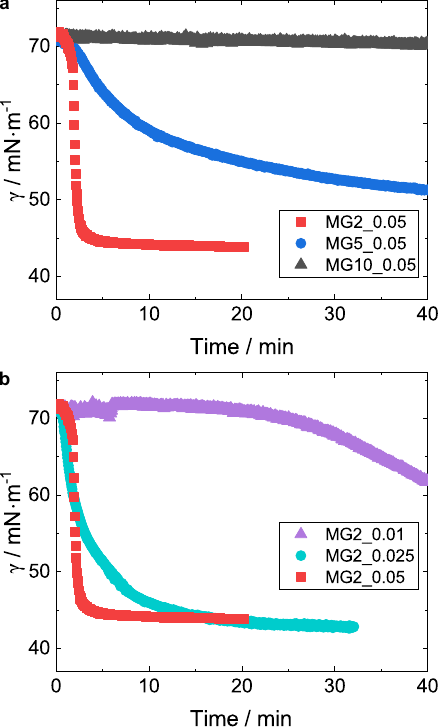}
  \caption{Kinetics of the surface tension $\gamma$ of a \mbox{MG-loaded} \mbox{air/water} interface as a function of a) \mbox{cross-linker} density and b) of MG concentration.}
  \label{fgr:IFT}
\end{figure}

Fig.~\ref{fgr:IFT} shows the kinetics of the surface tension $\gamma$ for a \mbox{MG-loaded} \mbox{air/water} interface, determined by pendant drop tensiometry, as a function of a) \mbox{cross-linker} density and b) MG concentration.
The decrease in $\gamma$ is faster for MGs with lower \mbox{cross-linker} density, again owing to their greater deformability and the higher abundance of dangling chains, thus, covering a larger surface area with a constant number of MGs.
Furthermore, the decrease in $\gamma(t)$ is accelerated by increasing the number of available MGs, \textit{i.e.} with increasing MG concentration \cite{Deshmukh2014-st, Stock2024-zy}. 

\subsection{Mobility in free-standing foam films}
\begin{figure}[htb]
\centering
  \includegraphics[height=5cm]{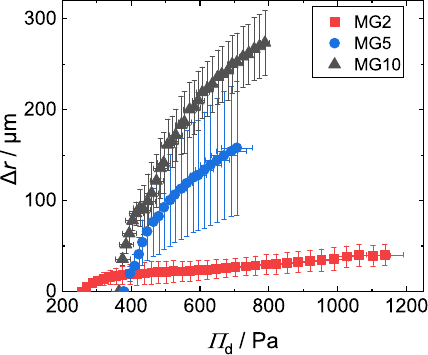}
  \caption{Feature displacement $\Delta r$ as a function of disjoining pressure $\varPi_\mathrm{d}$ in \mbox{free-standing} foam films stabilised by MG2, MG5 and MG10.}
  \label{fgr:Mobility}
\end{figure}

Next, the mobility in \mbox{free-standing} foam films is examined to determine whether the increase in $E_\mathrm{s}$ at the first elastic maximum with decreasing \mbox{cross-linker} density, observed for a single \mbox{air/water} interface in the Langmuir trough experiments, also corresponds to a greater foam film stiffness.
In this respect, the mobility in the foam films is quantified by the feature displacement $\Delta r$, which is shown in Fig.~\ref{fgr:Mobility} as a function of disjoining pressure $\varPi_\mathrm{d}$ for foam films stabilised by MG2, MG5 and MG10.
The displacement increases with increasing \mbox{cross-linker} density, indicating a low mobility in foam films stabilised by MG2, and intermediate mobility with MG5 and a high mobility with MG10.
The displacement curves end when the foam films rupture.

\subsection{Macroscopic foams}
After examining the interfacial properties of \mbox{MG-stabilised} planar \mbox{air/water} interfaces and single, \mbox{free-standing} foam films, the analysis is extended to macroscopic foams, representing the largest length scale of the \mbox{multi-scale} investigation.
The effect of \mbox{cross-linker} density and MG concentration on foam formation and stability are evaluated.
Foam formation is investigated from the start of the experiment until $t_\mathrm{0}$, where the foam volume of $V_\mathrm{foam} = \SI{100}{\milli\liter}$ has been generated.
The behaviour of the foam after $t_\mathrm{0}$ is examined to assess foam stability. 
\subsubsection{Foam formation}
\begin{figure*}[htb]
 \centering
 \includegraphics[height=10cm]{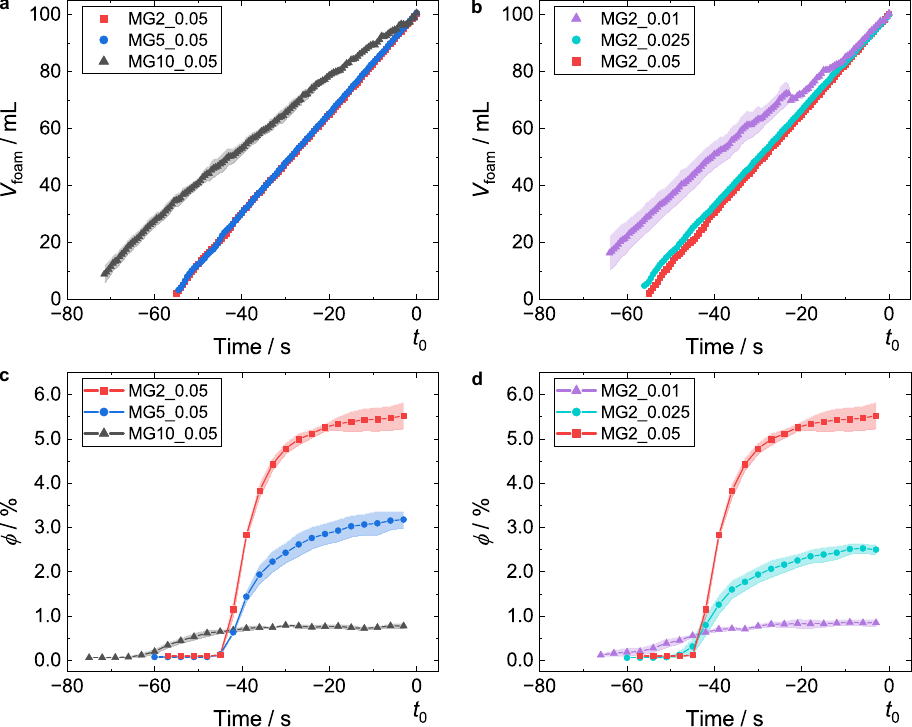}
 \caption{Time evolution of the foam volume $V_\mathrm{foam}$ during foam formation as a function of a) \mbox{cross-linker} density and b) MG concentration. Please note that in a) the curves for MG2 and MG5 lie on top of each other. Panels c) and d) show the time evolution of the liquid fraction $\phi$ in the foam under the same conditions. Since the individual repetitions may have differed in their foaming times, mean values and error bars were calculated only for time points available across all repetitions. Consequently, the curves start at the earliest time point for which all repetitions contributed to the mean values. In addition, some error bars are of the same order as the symbol size and may therefore not be visible.}
 \label{fgr:foamability_T_LF}
\end{figure*}

The time evolution of the foam volume $V_\mathrm{foam}$ and liquid fraction $\phi$ in the foam during foam formation is depicted in Fig.~\ref{fgr:foamability_T_LF}.
Fig.~\ref{fgr:foamability_T_LF}a) and b) show $V_\mathrm{foam}(t)$ as a function of \mbox{cross-linker} density and MG concentration, respectively.
The foam volume $V_\mathrm{foam}$ increases linearly in samples with low or intermediate \mbox{cross-linker} densities (MG2\_0.05 and MG5\_0.05) or intermediate and high MG concentrations (MG2\_0.025 and MG2\_0.05).
A \mbox{non-linear} increase occurs for the highest \mbox{cross-linker} density (MG10\_0.05) or lowest MG concentration (MG2\_0.01).
In addition, the curves are shifted along the time axis such that the point in time where the foam reaches its target volume of $V_\mathrm{foam} = \SI{100}{\milli\liter}$ is at \SI{0}{\second}, which corresponds to $t_\mathrm{0}$.
In Fig.~\ref{fgr:foamability_T_LF}a) the foam stabilised by MG10 requires more time to reach the target foam volume than the foams stabilised by MG2 and MG5.
Similarly, Fig.~\ref{fgr:foamability_T_LF}b) shows that the lowest MG concentration of \SI{0.01}{wt\percent} has a longer time period for foam formation than the higher MG concentrations.
Furthermore, a longer foam formation time is related to a \mbox{non-linear} increase in foam volume during foam formation. 

A time series of CCD1 photographs of the foam analyser tube acquired at selected times during foam formation for a representative MG at low concentration (MG2\_0.01) is shown in Fig.~S5 in the ESI\dag.
Here we observed that a \mbox{non-linear} increase in foam volume is accompanied by simultaneous foam collapse during foam formation.
In addition, another experimental parameter inducing a transition from a \mbox{non-linear} to linear increase in foam volume is the gas flow rate $Q$, presented in Fig.~S6 in the ESI\dag{}.  
At high $Q$ ($Q<\SI{20}{\milli\liter\per\minute}$), $V_\mathrm{foam}$ increases linearly.
However, if $Q$ is low ($Q=\SI{20}{\milli\liter\per\minute}$), $V_\mathrm{foam}$ increases in a \mbox{non-linear} manner, similar to MG10\_0.05 and MG2\_0.01 in Fig.~\ref{fgr:foamability_T_LF}.

During foam formation, not only the foam volume increases but also the liquid fraction $\phi$ in the foam, as shown in Fig.~\ref{fgr:foamability_T_LF}c) and d) as a function of \mbox{cross-linker} density and MG concentration, respectively.
The offset in time between the increase in $\phi$ compared to the increase of $V_\mathrm{foam}$ is attributed to the position of the electrode (E1), by which $\phi$ is determined, as it takes time for the rising foam to reach the electrode.
Similarly, because $\phi$ is measured at a fixed position, its value initially increases sharply, before reaching a plateau. 
The plateau value increases both with decreasing \mbox{cross-linker} density and with increasing MG concentration.
A magnified view of $\phi(t)$ of samples MG10\_0.05 and MG2\_0.01 is provided in Fig.~S7 in the ESI\dag, showing intermittent drops during the rise of $\phi(t)$.

\begin{figure*}[htb]
 \centering
 \includegraphics[height=13cm]{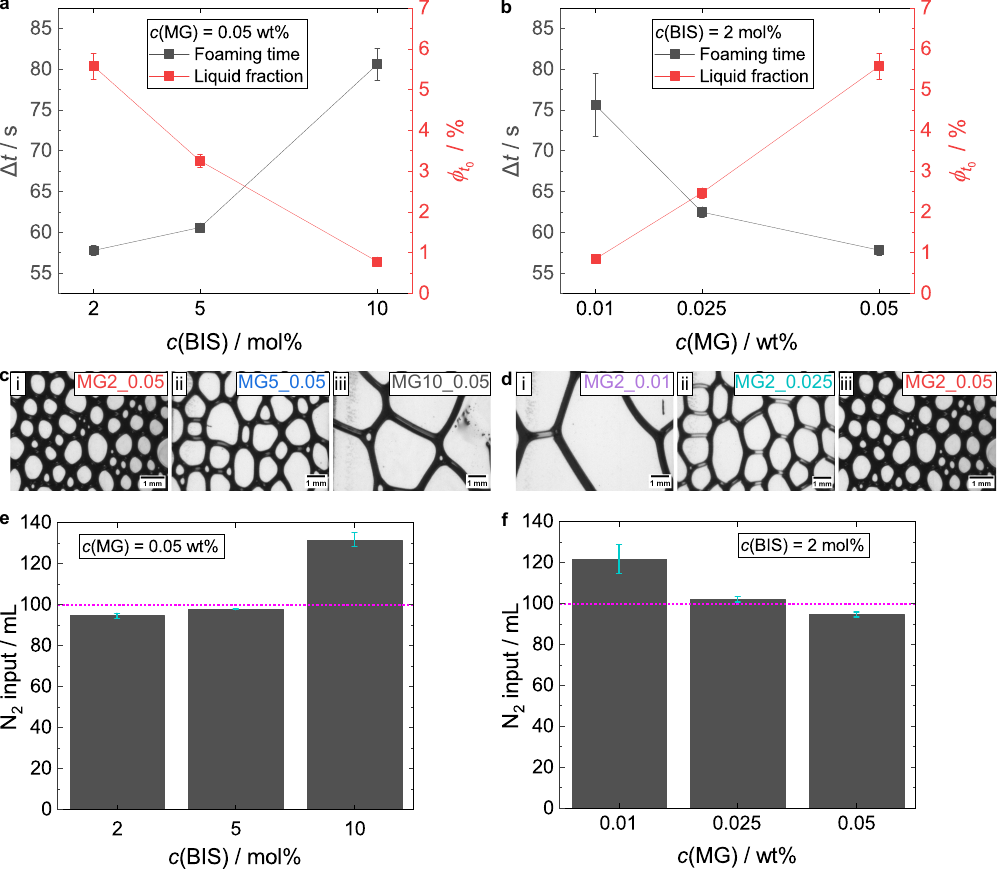}
 \caption{Time period $\Delta t$ required to generate a foam volume of $V_\mathrm{foam} = \SI{100}{\milli\liter}$ with a flow rate of $Q = \SI{100}{\milli\liter\per\minute}$, and liquid fraction at $t_0$ $\phi_{\mathrm{t_0}}$, shown as a function of a) \mbox{cross-linker} content and b) MG concentration. CCD2 photographs of foam bubbles at the tube wall, acquired at $t_\mathrm{0}$ are shown for c) varying cross-linker content: (i) MG2\_0.05, (ii) MG5\_0.05 and (iii) MG10\_0.05, and d) varying MG concentration: (i) MG2\_0.01, (ii) MG2\_0.025 and (iii) MG2\_0.05. Dark features correspond to liquid and bright features to gas. The scale bar represents \SI{1}{\milli\meter}; its absolute length might deviate slightly between images if the camera was repositioned and recalibrated between measurements. The \ce{N2} input volume required to generated $V_\mathrm{foam} = \SI{100}{\milli\liter}$ is shown as a function of e) \mbox{cross-linker} content and f) MG concentration. The theoretical maximum \ce{N2} input volume of \SI{100}{\milli\liter} is indicated by a pink dashed line in both plots.}
 \label{fgr:foamability_deltat_gas}
\end{figure*}

As stated above, depending on the \mbox{cross-linker} density and MG concentration, the foams require different time periods to reach the target foam volume of \SI{100}{\milli\liter}.
This time period can be denoted as $\Delta t$, and is a common parameter to quantify the foam forming capability of a liquid, also called foamability \cite{Boos2013-us}.
Fig.~\ref{fgr:foamability_deltat_gas} shows the $\Delta t$ values of the systems studied in this work as a function of a) \mbox{cross-linker} density and b) MG concentration. 
In addition, the liquid fraction at $t_\mathrm{0}$ $\phi_\mathrm{t_0}$  is given on the secondary ordinate.
Both an increasing \mbox{cross-linker} density and decreasing MG concentration increase $\Delta t$ and decrease $\phi_\mathrm{t_0}$.
In both cases, a smaller $\Delta t$ correlates with a larger liquid fraction in the foam at $t_\mathrm{0}$.

Moreover, a larger liquid fraction in the foam with decreasing \mbox{cross-linker} density or increasing MG concentration also becomes visible in the CCD2 photographs of the foams taken at $t_\mathrm{0}$, depicted in Fig.~\ref{fgr:foamability_deltat_gas}c) and d), respectively.
In Fig.~\ref{fgr:foamability_deltat_gas}c), the bubble size, corresponding to bright regions in the image, increases with increasing \mbox{cross-linker} density from c(i) to c(iii), whereas the liquid volume, corresponding to dark regions, decreases.
The opposite trend is visible in Fig.~\ref{fgr:foamability_deltat_gas}d)  with increasing MG concentration from d(i) to d(iii).
The Sauter diameter $D_\mathrm{s}$ of the foam bubbles at $t_\mathrm{0}$ are $D_\mathrm{s} \mathrm{(MG2\_0.05)}$=\SI{1226 \pm 155}{\micro\meter}, $D_\mathrm{s} \mathrm{(MG5\_0.05)}$=\SI{1894 \pm 171}{\micro\meter} and $D_\mathrm{s} \mathrm{(MG2\_0.025)}$=\SI{2160\pm 91}{\micro\meter}.
For samples MG10\_0.05 and MG2\_0.01 no statistically relevant value of $D_\mathrm{s}$ could be determined due to the small number of bubbles in the image section.

Fig.~\ref{fgr:foamability_deltat_gas} furthermore shows the volume of \ce{N2} (input gas) required to generate the target foam volume of $V_\mathrm{foam} = \SI{100}{\milli\liter}$ as a function of e) \mbox{cross-linker} density and f) MG concentration.
The required \ce{N2} volume increases with increasing \mbox{cross-linker} density and decreases with increasing MG concentration.
Moreover, the theoretical maximum  \ce{N2} volume of \SI{100}{\milli\liter}, indicated by a pink dashed line, is exceeded for the highest \mbox{cross-linker} density ($c$(BIS) = \SI{10}{mol\%}) and lowest MG concentration ($c$(MG) = \SI{0.01}{wt\%}). 

\label{Foam formation}

\subsubsection{Foam stability}
To assess foam stability, the drainage behaviour of the foams is investigated through the time evolution of the liquid fraction $\phi$ in the foam starting at $t_\mathrm{0}$, \textit{i.e.} after the target foam volume of \SI{100}{\milli\liter} has been generated.
We will first present the results on the impact of cross-linker density on foam stability, followed by the impact of MG concentration. 

\subsection*{Effect of \mbox{cross-linker} density}

\begin{figure*}[htb!]
 \centering
 \includegraphics[height=10cm]{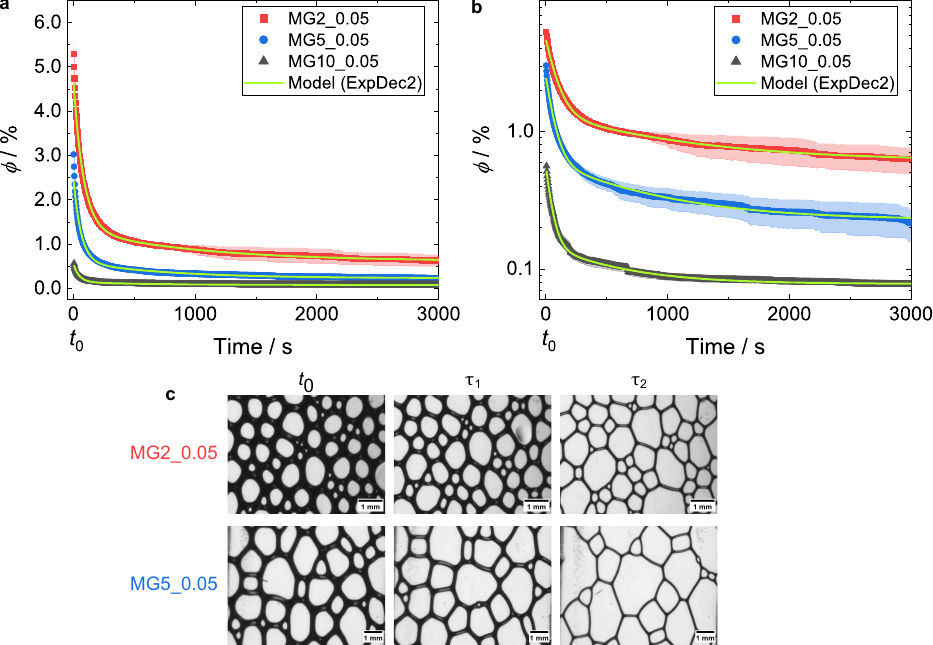}
 \caption{Time evolution of the liquid fraction $\phi$ in the foam as a function of \mbox{cross-linker} density starting at $t_\mathrm{0}$, plotted on a a) \mbox{lin-lin} scale and b) \mbox{log-lin} scale. In both graphs, symbols show the experimental data, while the solid green line represents the fit according to a \mbox{bi-exponential} decay model. c) Photographs of the foam bubbles at the characteristic times $t_\mathrm{0} = \SI{0}{\second}$, $\tau_\mathrm{1}$ and $\tau_\mathrm{2}$ of samples MG2\_0.05 ($\tau_\mathrm{1}=\SI{82}{\second}$; $\tau_\mathrm{2}=\SI{969}{\second}$) and MG5\_0.05 ($\tau_\mathrm{1}=\SI{61}{\second}$; $\tau_\mathrm{2}=\SI{757}{\second}$). In general, some error bars are of the same order as the symbol size and may therefore not be visible.}
 \label{fgr:stability_CL} 
\end{figure*}

Fig.~\ref{fgr:stability_CL} depicts $\phi(t)$ as a function of \mbox{cross-linker} density plotted on a) a \mbox{lin-lin} scale and b) a \mbox{log-lin} scale.
For the sake of completeness, a representation of $\phi(t)$ on a \mbox{log-log} scale is provided in Fig.~S8 in the ESI\dag.
In general, $\phi(t)$ decreases with time in all samples due to liquid drainage and collapse of the foam. 
Additionally, the initial $\phi$ values at $t_\mathrm{0}$ increase with decreasing \mbox{cross-linker} density as a consequence of the different individual foamabilities (see section ~\ref{Foam formation}).
Furthermore, in all samples $\phi(t)$ initially shows a rapid and pronounced decrease, followed by a more moderate decrease at longer times.
In this regard, the representation of $\phi(t)$ on a \mbox{log-lin} scale in Fig.~\ref{fgr:stability_CL}b) reveals that the decrease in $\phi(t)$ does not follow a \mbox{mono-exponential} decay model. 
Rather, $\phi(t)$ can be modelled by a \mbox{bi-exponential} decay according to $ y(x) = A_1 \exp\left[-\frac{x'}{\tau_1}\right]+A_2 \exp\left[-\frac{x'}{\tau_2}\right]+y_0$ (for further information see section 7 in the ESI\dag), with the respective fits indicated by green solid lines in Fig.~\ref{fgr:stability_CL}a) and b).
The \mbox{bi-exponential} decay model gives rise to two characteristic decay constants $\tau_\mathrm{1}$ and $\tau_\mathrm{2}$, which are listed in Tab.~\ref{tbl:bi-exp_dec}.
The other fit parameters are specified in Tab.~S1 in the ESI\dag.
This model suggests that two regimes govern the decrease in $\phi(t)$ \textit{i.e.} a fast decay regime with decay constant $\tau_\mathrm{1}$ followed by a slower decay regime with decay constant $\tau_\mathrm{2}$.
In general, both $\tau_\mathrm{1}$ and $\tau_\mathrm{2}$ decrease with increasing \mbox{cross-linker} density, implying faster decrease in $\phi(t)$ in both decay regimes with increasing \mbox{cross-linker} density.

\begin{figure*}[htb]
 \centering
 \includegraphics[height=5cm]{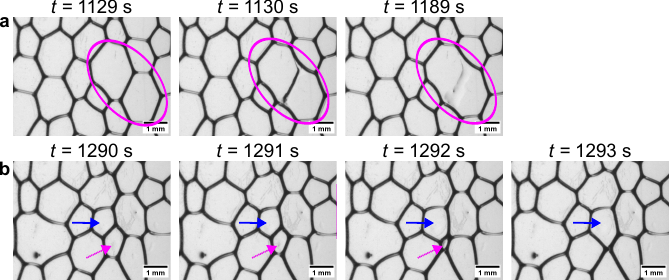}
 \caption{Representative examples of a) bubble coalesence and b) Ostwald ripening during foam collapse, visualised by photographs of the foam bubbles of a MG5\_0.05 foam recorded by the CCD2 camera at different times. The pink circle in a) is the region of interest highlighting bubble coalesence which occurs from $t=\SI{1129}{\second}$ to $t=\SI{1130}{\second}$. The image at $t=\SI{1189}{\second}$ is included to show the state after part of the liquid released during the coalesence has drained away, so that is becomes clearly visible that only one bubble remains within the region of interest. In b) the blue arrow and pink dashed arrow indicate two bubbles which increase (blue arrow) and decrease (pink dashed arrow) in size over time or may eventually disappear completely at $t=\SI{1293}{\second}$ as a result of Ostwald ripening.} \label{fgr:stability_rupture_coalesence}
\end{figure*}

\begin{table}[h]
\small
  \caption{Characteristic decay times $\tau_\mathrm{1}$ and $\tau_\mathrm{2}$ obtained as fit parameters from the bi-exponential decay model of the time dependent decrease in liquid fraction $\phi$ as a function of \mbox{cross-linker} density. Uncertainties represent the standard errors of the fit parameters}
  \label{tbl:example1}
  \begin{tabular*}{0.48\textwidth}{@{\extracolsep{\fill}}lll}
    \hline
    Foam & $\tau_1$ / \si{\second} & $\tau_2$ / \si{\second} \\
    \hline
    MG2\_0.05 & $81.6 \pm 0.5 $ & $969 \pm 13$ \\
    MG5\_0.05 & $60.9\pm 0.7$ & $757 \pm 13$ \\
    MG10\_0.05 & $47.1 \pm 0.7$ & $556\pm 6$ \\   
    \hline
    \label{tbl:bi-exp_dec}
  \end{tabular*}
\end{table}
To analyse the reason for a \mbox{two-step} decay we investigated the bubble development over time.
Hence, Fig.~\ref{fgr:stability_CL}c) depicts bubble images of samples MG2\_0.05 and MG5\_0.05 captured at $t_\mathrm{0}$, $\tau_\mathrm{1}$ and $\tau_\mathrm{2}$.
The image series show that with time the liquid in the foam, represented by the dark parts in the image, decreases strongly, the Plateau borders separating adjacent bubbles become thinner, and that the initial bubble network at $t_\mathrm{0}$ becomes disrupted.
Disruption of the bubble network is likely caused by bubble coalescence, and Ostwald ripening, where larger bubbles grow at the expense of smaller bubbles.
Representative examples of these processes are visualised in Fig.~\ref{fgr:stability_rupture_coalesence} by photographs of foam bubbles at different times.
In Fig.~\ref{fgr:stability_rupture_coalesence}a) two bubbles are visible within the pink circle at $t=\SI{1129}{\second}$, whereas only one bubble remains at $t=\SI{1130}{\second}$, indicating bubble coalesence.
In addition, the image at $t=\SI{1189}{\second}$ is included to show the state after part of the liquid released during the coalesence has drained away, so that is becomes clearly visible that only one bubble remains within the region of interest (pink circle).
Fig.~\ref{fgr:stability_rupture_coalesence}b) shows the time the evolution of two bubbles marked by a blue arrow and pink dashed arrow. 
At $t=\SI{1290}{\second}$ both bubbles are relatively similar in size. 
Over time the bubble marked by the blue arrow increases in size, whereas the bubble marked by the pink dashed arrow decreases in size and completely disappears at $t=\SI{1293}{\second}$, visualising the process of Ostwald ripening.

\subsection*{Effect of MG concentration}
\begin{figure*}[htb]
 \centering
 \includegraphics[height=5cm]{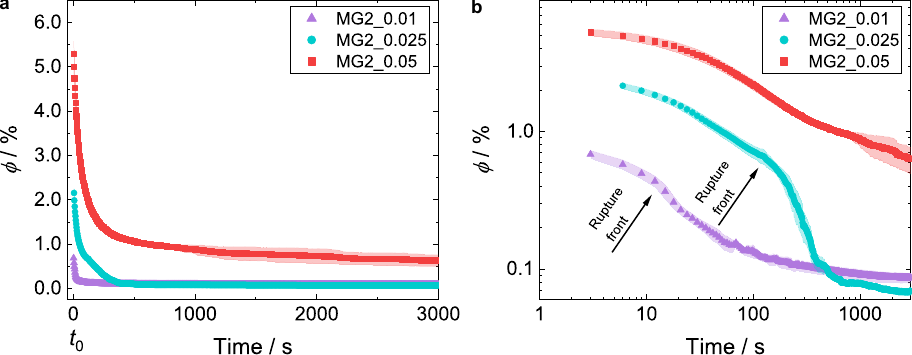}
 \caption{Time evolution of the liquid fraction $\phi$ in the foam as a function of MG concentration starting at $t_\mathrm{0}$, plotted on a a) \mbox{lin-lin} scale and b) \mbox{log-log} scale. In addition, the onset of the rupture front occurring at $\approx\SI{15}{\second}$ for MG2\_0.01 and $\approx\SI{120}{\second}$ for MG2\_0.025 is indicated by an arrow in b).}
 \label{fgr:stability_CONC}
\end{figure*}

Fig.~\ref{fgr:stability_CONC} depicts $\phi(t)$ as a function of MG concentration plotted on a) a \mbox{lin-lin} scale and b) a \mbox{log-log} scale.
For the sake of completeness, a representation of $\phi(t)$ on a \mbox{log-lin} scale is provided in Fig.~S8 in the ESI\dag.
Similarly, the initial $\phi$ values at $t_\mathrm{0}$ increase with increasing MG concentration as a result of the different individual foamabilities of the foaming dispersions (see section ~\ref{Foam formation}).
Apart from this, it becomes apparent that the decay behavior of $\phi(t)$ changes significantly when the MG concentration is decreased from \SI{0.05}{wt\percent}, to \SI{0.025}{wt\percent} or \SI{0.01}{wt\percent}.
In this regard, the decrease in $\phi(t)$ of samples MG2\_0.025 and MG2\_0.01, shown on a \mbox{log-log} scale in Fig.~\ref{fgr:stability_CONC}b), reveals the presence of a foam rupture front, which induces an abrupt drop in $\phi$ occurring at $\approx\SI{15}{\second}$ for MG2\_0.01 and $\approx\SI{120}{\second}$ for MG2\_0.025.
A rupture front arises from total foam collapse rather than drainage \cite{Boos2012-bw, Mikhailovskaya2022-ts}.
Because the electrodes by which $\phi$ is determined (E1) are installed at a fixed position, and they are no longer covered by foam once the propagating rupture front has passed, the remaining signal measured is that of the wetting layer on the electrodes \cite{Mikhailovskaya2022-ts}.

\section{Discussion}
Both the MG \mbox{cross-linker} density as well as the MG concentration significantly influence the foamability and stability of macroscopic foams.
In the following discussion, the macroscopic foam characteristics are correlated with the properties of single \mbox{air/water} interfaces (Langmuir compression and pendant drop tensiometry), as well as those of two \mbox{air/water} interfaces in close proximity to each other (\mbox{free-standing} foam films).

\subsection{Foam formation}
The foamability of the dispersions was enhanced both by decreasing the \mbox{cross-linker} density in the stabilising MGs and by increasing the MG concentration.
This was expressed by shorter foam formation times $\Delta t$, smaller bubble radii and higher liquid fractions at $t_\mathrm{0}$ with decreasing  \mbox{cross-linker} density or increasing or MG concentration.

Furthermore, the foam volume $V_\mathrm{foam}$ of samples MG2\_0.05, MG5\_0.05 and MG2\_0.025 increased linearly with time, thus, implying a direct conversion of the input energy into formation of foam.
In contrast, the \mbox{non-linear} increase of $V_\mathrm{foam}$ in samples MG10\_0.05 and MG2\_0.01 indicates processes counteracting foam growth due to partial foam collapse during foam formation; this is also reported in literature for foams stabilised by surfactants \cite{Boos2012-bw} and proteins \cite{Saint-Jalmes2005-tf}, respectively.
Similar observations were made at low flow rates, where the foam is also slowly formed.
Determination of $V_\mathrm{foam}$ is based on measurement of the foam height.
When the foam starts to partially collapse during formation, the rate at which the foam height increases slows down, and consequently, the foam volume increases in a \mbox{non-linear} way.
Partial foam collapse during foam formation also explains the increased \ce{N2} volume required for the generation of a target foam volume of $V_\mathrm{foam} = \SI{100}{\milli\liter}$ in samples MG10\_0.05 and MG2\_0.01.
The theoretical maximum of \SI{100}{\milli\liter} of \ce{N2} for generation of $V_\mathrm{foam} = \SI{100}{\milli\liter}$ is exceeded because not all the supplied \ce{N2} remains trapped in the foam; instead, part of the gas is released again as a result of partial foam collapse.
In addition, the intermittent drops of the liquid fraction $\phi$ in the foam observed during foam formation in samples MG10\_0.05 and MG2\_0.01 result from bubble coalescence and the associated abrupt release of larger liquid volumes, and support the interpretation of foam collapse during foam formation.
Therefore, the simultaneous occurrence of foam formation and collapse contributes to a significantly increased $\Delta t$ of samples MG10\_0.05 and MG2\_0.01.

In the next step, this macroscopic foam behavior is put into relation with the results obtained from the pendant drop tensiometry investigating the kinetics of the surface tension $\gamma$ of a single \mbox{air/water} interface. 
In this respect, a correlation between a higher foamability and faster decrease in $\gamma$ is known from \mbox{surfactant-stabilised} foams \cite{Prins1992,Pugh1996, Engels2008-wr}.
Similarly, we also correlate a higher foamability to a faster decrease in $\gamma$, achieved by decreasing \mbox{cross-linker} density and increasing MG concentration.
The lower cross-linked MGs accelerate the decrease in $\gamma$ because they have smaller elastic moduli $E_\mathrm{MG}$, are therefore softer and, hence, capable of greater deformation when adsorbed at the \mbox{air/water} interface allowing for coverage of a larger surface area with the same number of MGs in contrast to the stiffer MGs (higher $E_\mathrm{MG}$) with higher \mbox{cross-linker} density.
Thus, an elevated decrease in $\gamma(t)$ correlates with an increased foamability because the lower \mbox{cross-linked} MGs are capable of stabilising the freshly generated \mbox{air/water} interface faster with the same energy input.
The result is the stabilisation of a larger internal foam surface area and, therefore, a foam with smaller bubbles and higher liquid fraction at $t_\mathrm{0}$.
A similar effect is achieved by increasing the absolute number of MG available in the dispersion by increasing the MG concentration, because more MGs adsorb at the freshly generated \mbox{air/water} interface in the same period of time.

Still, interestingly, the decrease in $\gamma(t)$ takes place on much longer time scales compared to the time scales of foam formation. 
What is more, $\gamma(t)$ of samples MG10\_0.05 and MG2\_0.01 did not decrease significantly within the time frame investigated ($\SI{40}{\minute}$).
Nevertheless, it was possible to generate foams from all these dispersions.
This is, because contrary to the measurement of $\gamma(t)$ by pendant drop tensiometry, in which the MG adsorption to the \mbox{air/water} interface occurs spontaneously, the macroscopic foams are generated with a higher energy input by sparing gas through the MG dispersion, which likely accelerates the MGs adsorption to newly generated \mbox{air/water} interfaces by increasing the frequency of \mbox{particle-interface} encounters.

Apart from the different MG adsorption kinetics, another effect influencing the foamability could be depletion.
Here depletion refers to the gradual reduction of the stabilising agent in the liquid bulk phase during foam formation and can influence the foam properties, especially if the concentration of the stabilising agents is low and a large internal interface is generated \cite{Boos2012-bw}.
This effect could be particularly relevant at the lowest MG concentration of \SI{0.01}{wt\percent} and another reason for the partial foam collapse during foam formation in this sample.
However, a relatively large bulk liquid volume of \SI{60}{\milli\liter} is available for the generation of \SI{100}{\milli\liter} of foam, which increases the absolute number of MGs present in the system.

In general, the foamability increased with increasing MG concentration which is in line with the work of Kühnhammer \textit{et al.} \cite{D2SM01021F}, where \mbox{MG-stabilised} foams were generated by the Bartsch method.
An increased foamability with increasing concentration of \mbox{stabilising agent} is also known for other systems \textit{e.g.} \mbox{surfactant-stabilised} \cite{Stubenrauch2009-wk}, \mbox{particle-stabilised} \cite{Tyowua2020} and \mbox{protein-stabilised}  macroscopic foams \cite{Saint-Jalmes2005-tf}.

\subsection{Foam stability}
The Langmuir compression experiments showed that the surface elastic modulus $E_\mathrm{s}$ at the first elastic maximum increases with decreasing \mbox{cross-linker} density, indicating that interfacial monolayers formed by less \mbox{cross-linked} MGs exhibit greater resistance to compression in this regime.
We propose that this may result from steric repulsion between the dangling polymer chains of the MGs adsorbed at the interface, arising from a reduction in their conformational entropy upon compression.
The steric repulsion, and hence the resistance to compression, may increase with decreasing \mbox{cross-linker} density because the less \mbox{cross-linked} MGs exhibit a more pronounced \mbox{core-shell} character, as observed in the AFM scans, associated with a higher abundance of dangling polymer chains.

To investigate whether the increase in $E_\mathrm{s}$ at the first elastic maximum with decreasing \mbox{cross-linker} density, also corresponds to a greater foam film stiffness, the mobility in \mbox{free-standing} foam films was determined using feature tracking.
The mobility in the foam films decreased with decreasing \mbox{cross-linker} density.
This may be attributed to greater interpenetration of the dangling polymer chains and formation of intermolecular hydrogen bonds, which could  increase the mechanical strength of the \mbox{air/water} interface and thereby reduce its mobility with decreasing \mbox{cross-linker} density.

With respect to the stability of macroscopic foams stabilised by MGs with varying \mbox{cross-linker} density the \mbox{time-dependent} decrease in liquid fraction $\phi$ followed a \mbox{bi-exponential} decay model, indicating two decay regimes occurring on different time scales with the decay constants $\tau_\mathrm{1}$ and $\tau_\mathrm{2}$.
Such a \mbox{two-step} decay of $\phi(t)$ has also been reported for other systems \textit{e.g.} surfactant-stabilised foam \cite{Lamolinairie2022-ew}.
In the first decay regime $\phi(t)$ decreased strongly and rapidly, followed by a slower, more moderate decrease in the second decay regime.
In both decay regimes the decrease in $\phi(t)$ was accelerated with increasing \mbox{cross-linker} density.
The processes associated with the decrease in $\phi(t)$ are liquid drainage, bubble coalescence and Ostwald ripening.
Therefore, we assume that a faster decrease in $\phi(t)$ reflects a faster progression and higher frequency of these destabilisation processes.
In the following the effect of \mbox{cross-linker} density on these macroscopic foam destabilisation processes is discussed in relation to the findings obtained at different length scales.

Regarding liquid drainage, the mobility of the \mbox{air/water} interface determines the boundary condition for the liquid flow and, thus, where most of the viscous dissipation during drainage occurs.
In this respect, an immobile \mbox{air/water} interface is associated with a \mbox{stress-carrying}/\mbox{no-slip} boundary condition resulting in a \mbox{Poiseuille-like} flow, where the viscous dissipation is dominated by Plateau borders.
In contrast, a mobile \mbox{air/water} interface corresponds to a \mbox{stress-free}/slip boundary condition, associated with a \mbox{Plug-like} flow.
In this case, the viscous dissipation is dominated by the nodes \cite{Durand2002-jz,Koehler2000-rf, Koehler2004-lc,Stone2003-us, Saint-Jalmes2004-ag, Cervantes-Martinez2005-bj}.
Thus, a slower drainage with decreasing \mbox{cross-linker} density may, in part, be associated with a lower interfacial mobility and a higher surface elastic modulus, increasing the interfacial shear resistance to liquid flow.
This interpretation is consistent with the findings for both \mbox{free-standing} foam films and the single \mbox{air/water} interface.

In addition, a lower interfacial mobility and higher surface elastic modulus could increase the foam film resistivity towards external lateral disturbances such as external vibrations, thermal fluctuations and temperature gradients, thus, reducing bubble coalescence \cite{Pugh2016-bp}.

Thirdly, during Ostwald ripening, larger bubbles grow at the expense of smaller bubbles. 
Because the MGs adsorb irreversibly at the \mbox{air/water} interface, and, thus, at the bubble surface, the MG monolayer becomes increasingly compressed as the bubble shrinks.
Since the Langmuir trough experiments at the single \mbox{air/water} interface demonstrated a greater resistance to lateral compression of monolayers formed by less \mbox{cross-linked} MGs, bubble shrinkage may be slowed down or counteracted with decreasing \mbox{cross-linker} density.
Conversely, bubble growth leads to expansion of the \mbox{MG-covered} \mbox{air/water} interface.
In this respect, Razavi \textit{et al.} \cite{Razavi2025-yu}, who investigated \mbox{MG-stabilised} \mbox{nozzle-free} water jets, showed that upon stretching of the interface lower \mbox{cross-linked} MGs (\SI{2}{\mol\percent} BIS) remained interconnected throughout the strain due to substantial chain interpenetration and high MG deformability.
In contrast, the higher \mbox{cross-linked} MGs (\SI{10}{\mol\percent} BIS) disentangled from each other more readily, causing the surface tension to rapidly approach that of the bare \mbox{air/water} interface.
Although \mbox{nozzle-free} jetting occurs on much shorter timescales than bubble growth, a greater interpenetration of dangling polymer chains and formation of intermolecular hydrogen bonds with decreasing \mbox{cross-linker} density could enhance the mechanical stability of the interfacial layer.

With regard to a varying MG concentration, only sample MG2\_0.05 exhibited the two decay regimes of $\phi(t)$ discussed above.
In contrast, for lower MG concentrations (MG2\_0.025 and MG2\_0.01) no second decay regime could be observed.
Instead drainage was interrupted by the onset of a sudden rupture front associated with foam collapse, often reported for aqueous foams \cite{Monin2000, Stubenrauch2009-wk, Boos2013-us, Mikhailovskaya2022-ts}.
The rupture front appeared earlier for sample MG2\_0.01 than for MG2\_0.025, suggesting lower foam stability of MG2\_0.01.
With increasing MG concentration, a larger internal area can be generated much faster, which leads to smaller and more bubbles and more stable foams.
In contrast, the bubble size increased with decreasing MG concentration.
This might be related to a lower surface coverage at low MG concentrations possibly caused by depletion of the MGs from the liquid bulk phase during foam formation.
A low surface coverage leaves less capacity to accommodate interfacial distortions or destabilisation processes.
Consequently, the ability to maintain a low interfacial tension may be reduced, promoting foam destabilisation and decreasing the overall foam stability.

Finally, the different foamabilities resulted in different foam morphologies.
The foams contained a larger number of smaller bubbles with decreasing \mbox{cross-linker} density and increasing MG concentration.
This finer foam structure could also contribute to a higher mechanical stability of the macroscopic foam as a whole.

\section*{Conclusion}
In this work, aqueous foams stabilised by PNIPAM microgels were investigated in a \mbox{multi-scale} approach to understand the interplay between single \mbox{air/water} interfaces, individual \mbox{free-standing} foam films and macroscopic foams and to examine the effect of \mbox{cross-linker} density and MG concentration.

In this context, the mobility in \mbox{MG-stabilised} \mbox{free-standing} foam films was, to our knowledge, analysed for the first time using feature tracking, which showed a decreasing mobility with decreasing \mbox{cross-linker} density.
AFM scans of the MGs revealed a more pronounced \mbox{core-shell} structure with decreasing \mbox{cross-linker} density, associated with a higher abundance of dangling polymer chains.
Thus, a greater interpenetration of the dangling polymer chains and formation of intermolecular hydrogen bonds could be a possible reason for the decreased mobility in the foam films.
In addition, a decreasing foam film mobility was interpreted as an indicator for greater stiffness and higher mechanical stability of the foam films.
This finding is consistent with a higher surface elastic modulus of lower \mbox{cross-linked} MGs observed at the first elastic maximum, \textit{i.e.} in the regime of spontaneous adsorption, by Langmuir compression experiments on a single \mbox{air/water} interface.
Furthermore, macroscopic foams were generated by sparging gas through a porous glass frit and the MG dispersions, respectively, investigating the foamability and foam stability.
A lower \mbox{cross-linker} density and higher MG concentration enhanced the foamability and generated foams with smaller bubbles and higher liquid fractions.
This was explained by faster stabilisation of the freshly generated \mbox{air/water} interfaces during macroscopic foam formation and correlated with a higher surface activity of MG dispersions with decreasing \mbox{cross-linker} density and increasing MG concentration observed at the single \mbox{air/water} interface by pendant drop tensiometry.
Macroscopic foam stability increased with decreasing \mbox{cross-linker} density, consistent with a higher surface elastic modulus of the single \mbox{air/water} interface and decreasing mobility in the foam films, possibly slowing down foam destabilisation processes such as liquid drainage, bubble coalescence and Ostwald ripening. 
Macroscopic foam stability also increased with increasing MG concentration, associated with a higher surface coverage allowing for greater resistance of the \mbox{air/water} interfaces to external disturbances. 
In general, the most stable foam was formed by the lowest \mbox{cross-linker} density and highest MG concentration.

To conclude, the results are consistent and in good agreement across all the length scales of \mbox{MG-stabilised} \mbox{air/water} interfaces, demonstrating a strong interplay between the differently sized building blocks of aqueous foams.
In addition, the individual interfacial properties are strongly linked to the characteristics of the stabilsing MGs.
These findings may be applicable to other foam systems, such as \mbox{protein-stabilised} foams, and even other interfaces such as those in emulsions. 
In addition, the temperature responsiveness of the PNIPAM MGs has the potential to create switchable foams that can be destabilised on demand. 
Therefore, a \mbox{follow-up} study investigating the influence of temperature on the \mbox{MG-stabilised} \mbox{air/water} interfaces at different length scales could enhance the understanding of \mbox{temperature-induced} foam destabilisation.

\section*{Author contributions}
Conceptualisation, J.Z., R.v.K.; formal analysis J.Z., L.M., K.G., C.S., H.R.; investigation, J.Z., L.M., G.B., V.H.;  resources R.v.K.; visualisation J.Z.; writing -- original draft, J.Z., writing -- review and editing, J.Z. L.M., K.G., C.S., H.R., R.v.K.; supervision, R.v.K.; funding acquisition, R.v.K., project administration R.v.K.

\section*{Conflicts of interest}
There are no conflicts to declare.

\section*{Data availability}
We hereby confirm that the data that support the findings of this study are openly available in the repository of the Technical University Darmstadt (TUDatalib) accessible via the link: \href{https://doi.org/10.48328/tudatalib-2302}{10.48328/tudatalib-2302}

\section*{Acknowledgements}

% The acknowledgements come at the end of an article after the conclusions and before the notes and references. 
K.G. is thankful for the funding by the DFG (German Research Council) within the project 395854042. C.S. is thankful for the funding by the DFG within the project KL1165/38-1, SPP 2494. H.R. gratefully acknowledges a Humboldt Research Fellowship from the Alexander von Humboldt Stiftung.

%%%END OF MAIN TEXT%%%


\balance

\renewcommand\refname{References}

%%%REFERENCES%%%
\bibliography{rsc} %You need to replace "rsc" on this line with the name of your .bib file
\bibliographystyle{rsc} %the RSC's .bst file
\end{document}

\maketitle

% \noindent
% \emph{All relevant data and code required to reproduce the analyses presented are readily available on Zenodo at} \url{https://doi.org/}.

\tableofcontents

\newpage
\section{Microgel height determined by AFM}

\begin{figure}[htb!]
    \centering    \includegraphics[height=10cm]{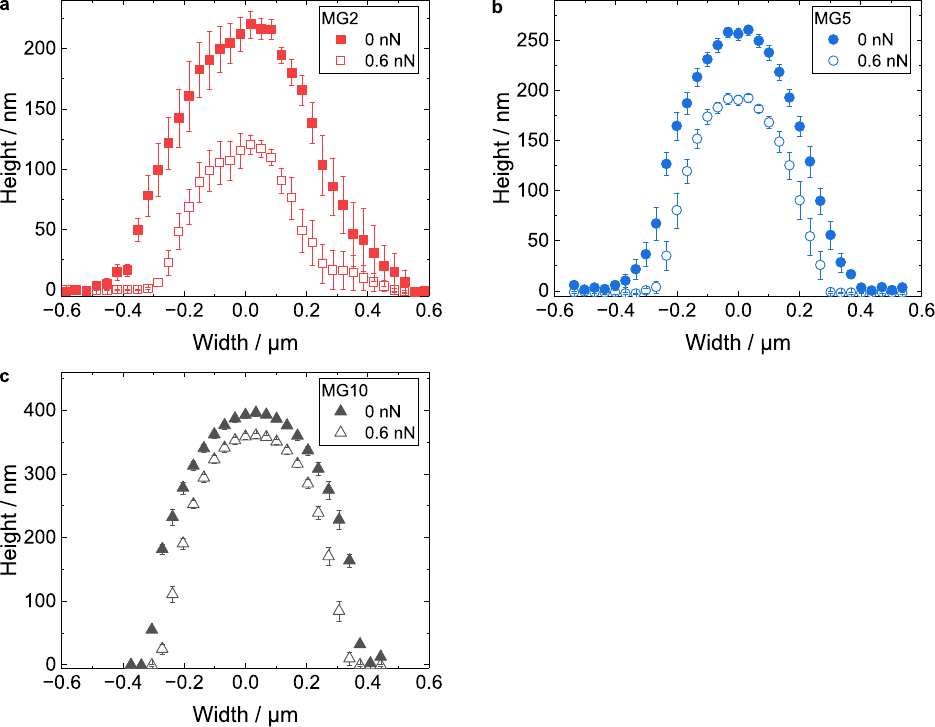}
    \caption{MG height determined by AFM in \ce{H2O} at \SI{20}{\celsius} without indentation (\SI{0}{\nano\newton}) and when indented with a force of \SI{0.6}{\nano\newton}, the maximum indentation force for the fit of the Hertz model, of a) MG2, b) MG5 and c) MG10.}
    \label{fgr:Height_RFH}
\end{figure}

In Fig.~\ref{fgr:Height_RFH} the \mbox{cross-sections} of the MG heights determined by atomic force microscopy (AFM) in \ce{H2O} at \SI{20}{\celsius} without indentation (\SI{0}{\nano\newton}) and when indented with a force of \SI{0.6}{nN} of microgels a) MG2, b) MG5 and c) MG10 are presented.
The MG height without indentation (0 nN) is extracted from the contact point offset, defined as the vertical distance between the substrate and MG contact point prior to indentation.
The reference force height corresponding to the MG height at a specified applied force, in this case \SI{0.6}{nN}, provides a measure of the MG height under indentation.
The indentation force of \SI{0.6}{nN} corresponds to the maximum indentation force for the fit of the Hertz model for determination of the MGs elastic modulus.
With increasing \mbox{cross-linker} densiy the MG height increases and the indentation depth decreases.

\newpage
\section{Image skeletonisation for determination of the bubble size}

\begin{figure} [htb]
\centering
  \includegraphics[height=7cm]{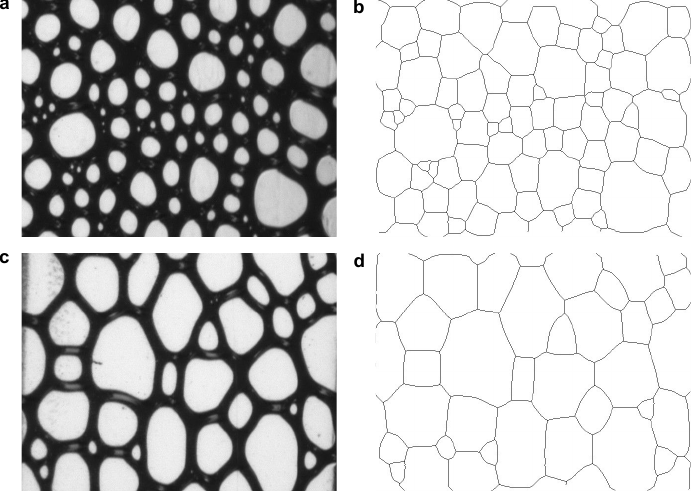}
  \caption{Representative examples of original images obtained by the CCD2 camera of the foam analyser device are shown in panels a) and c), together with the corresponding skeletonised versions shown in panels b) and d), which were used for subsequent processing with the CSA software to determine the Sauter bubble diameter.}
  \label{fgr:skeleton}
\end{figure}

Representative examples of the \mbox{image-processing} procedure for subsequent evaluation of the Sauter diameter of the foam bubbles using the Cell Size Analysis (CSA) software are presented in Fig.\ref{fgr:skeleton}.
Panels a) and c) show the original images acquired with the CCD2 camera of the foam analyser device.
These images were subjected to a skeletonisation procedure in ImageJ, in which the dark surface Plateau borders in the previously binarised image were reduced to \mbox{single-pixel} width, as shown in panels b) and d).
The skeletonised images were then analysed with the CSA software provided by TECLIS Scientific (Civrieux d’Azergues, France) to determine the Sauter diameter.

\newpage
\section{Photographs of the foam analyser tube during foam collapse}

\begin{figure} [htbp!]
\centering
  \includegraphics[height=1.15\textwidth]{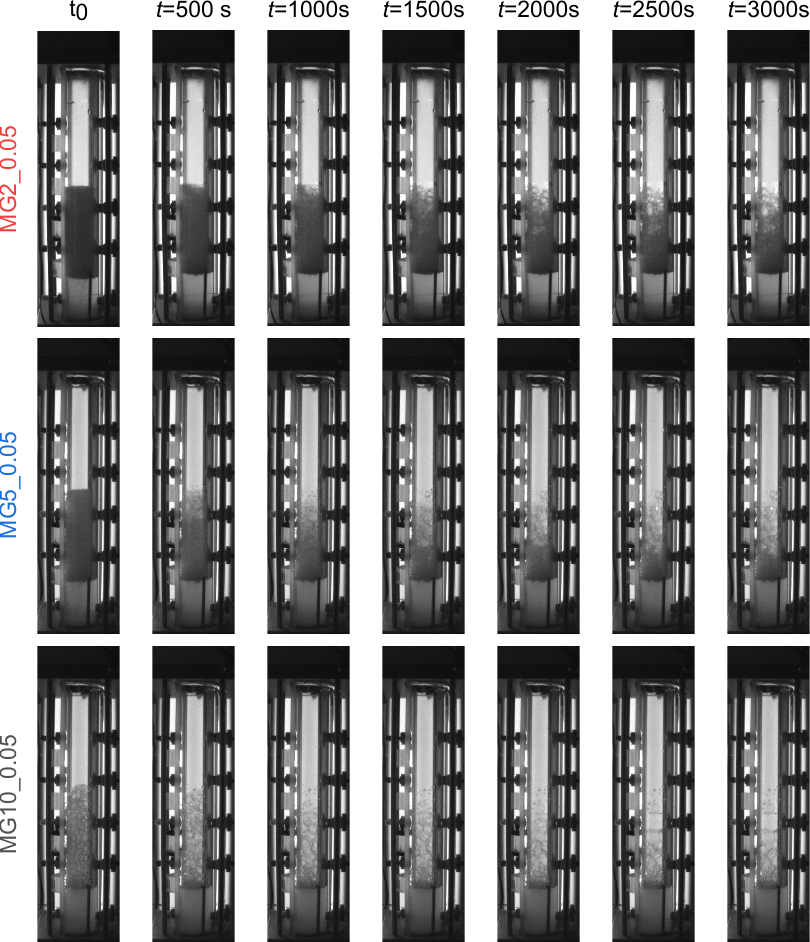}
  \caption{Representative CCD1 photographs acquired by the foam analyser device at selected times to visualise foam collapse as a function of increasing \mbox{cross-linker} density in the stabilising MGs for samples MG2\_0.05, MG5\_0.05 and MG10\_0.05.}
  \label{fgr:tube_stab_CL}
\end{figure}

\begin{figure} [htbp!]
\centering
  \includegraphics[height=1.15\textwidth]{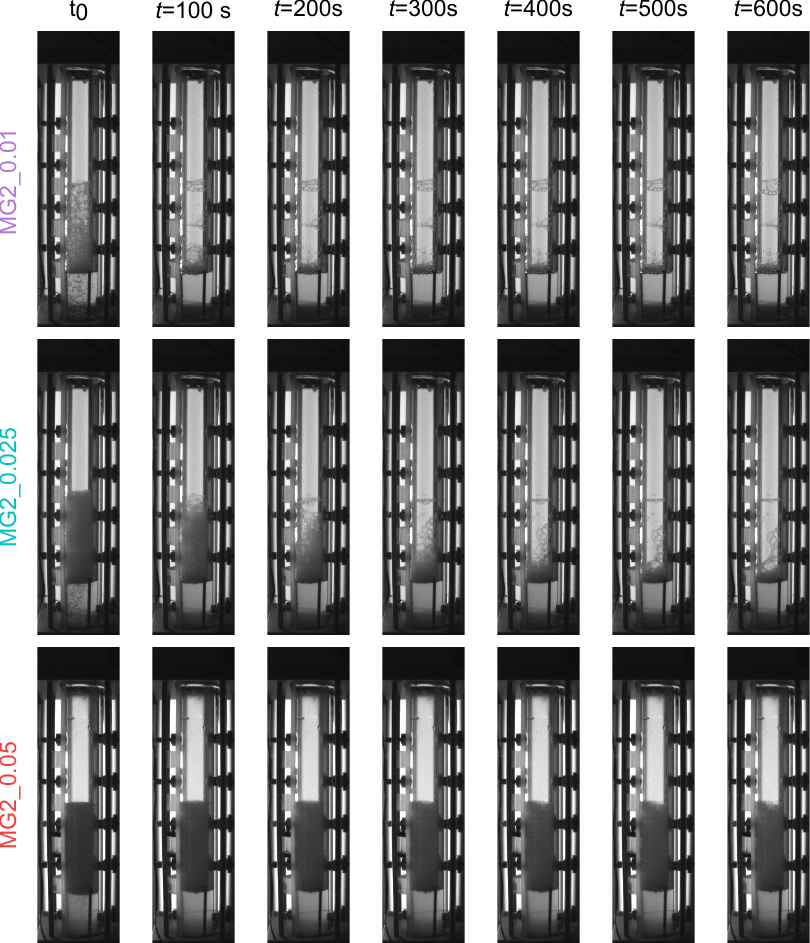}
  \caption{Representative CCD1 photographs acquired by the foam analyser device at selected times to visualise foam collapse as a function of increasing MG concentration for samples MG2\_0.01, MG2\_0.025 and MG2\_0.05.}
  \label{fgr:tube_stab_CONC}
\end{figure}

The time evolution of the foam collapse was captured by photographs of the foam analyser tube acquired with the CCD1 camera.
Representative images at selected times are shown as a function of \mbox{cross-linker} density in Fig.~\ref{fgr:tube_stab_CL} and as a function of MG concentration in Fig.~\ref{fgr:tube_stab_CONC}.

Both figures show that the foams at $t_\mathrm{0}$ appear brighter with increasing \mbox{cross-linker} density and decreasing MG concentration.
Since the foam analyser tube is positioned between a bright lamp homogeneously illuminating the tube and the CCD1 camera, the photographs show the foams in transmission.
Thus, the increased apparent brightness is consistent with larger bubble sizes and lower liquid fractions in the foam, which reduce the attenuation of the transmitted light.

Over time, the foams increase in apparent optical brightness as the decreasing liquid fraction in the foam, bubble growth and foam collapse reduce the attenuation of the transmitted light.

As a result, the foam of sample MG10\_0.05 appears very bright at longer experimental times, making it more difficult to distinguish by eye.
In this case this is particularly pronounced because this sample already exhibited relatively large bubbles and a low liquid fraction at $t_\mathrm{0}$ compared to the foams stabilised by MGs with lower \mbox{cross-linker} density.
However, upon closer inspection, a bright residual foam column is still visible in the lower part of the tube, \textit{e.g.} at $t=\SI{3000}{\second}$.

Because the foams are less stable with decreasing MG concentration, the photographs in Fig.~\ref{fgr:tube_stab_CONC} are shown only up to $t=\SI{600}{\second}$.
The photographs reveal, that MG2\_0.01 and MG2\_0.025 have almost completely collapsed after $t=\SI{100}{\second}$ and $t=\SI{500}{\second}$ respectively.

In general, the photographs demonstrate that, over time, the foams in all samples do not collapse with a \mbox{well-defined} \mbox{gas/foam} interface.
Instead, foam fragments may remain attached to the tube walls, which impedes the accurate determination of the foam volume during foam collapse.

\newpage
\section{Photographs of the foam analyser tube during foam formation}
\begin{figure} [htbp!]
\centering
  \includegraphics[height=7cm]{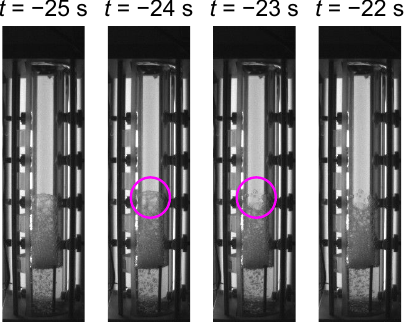}
  \caption{Representative CCD1 photographs of the foam analyser tube acquired at selected times during foam formation for a representative sample of MG2\_0.01. The pink circle highlights a region in which foam collapse occurs during foam formation.}
  \label{fgr:tube_form}
\end{figure}

Fig.~\ref{fgr:tube_form} shows a time series of CCD1 photographs of the foam analyser tube acquired with the CCD1 camera at selected times during foam formation for a representative MG2\_0.01 sample.
The pink circle highlights a region in which parts of the foam collapse during the foaming process.

\newpage
\section{Influence of flow rate $Q$ on foam formation}
\begin{figure} [htb]
\centering
  \includegraphics[height=5cm]{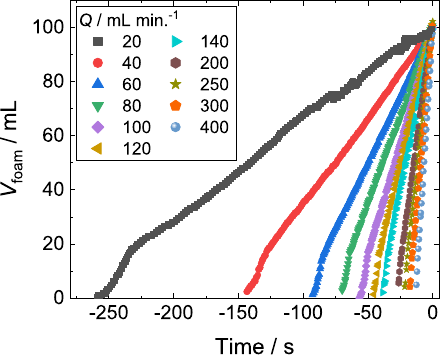}
  \caption{Time evolution of the foam volume during foam formation as a function of flow rate $Q$.}
  \label{fgr:Q_Formation}
\end{figure}

In Fig.~\ref{fgr:Q_Formation} the time evolution of the foam volume $V_\mathrm{foam}$ during foam formation is shown as a function of flow rate $Q$.
The time required for generation of a foam volume of $V_\mathrm{foam}=\SI{100}{\milli\liter}$ increases with decreasing $Q$.
At flow rates $Q<\SI{20}{\milli\liter\per\minute}$, $V_\mathrm{foam}$ increases linearly implying a direct transition of input energy into the formation of foam.
At a flow rate of  $Q=\SI{20}{\milli\liter\per\minute}$, $V_\mathrm{foam}$ increases \mbox{non-linearly}, indicating foam collapse during foam formation, because foam formation is so slow that the formed foam already collapses.

Please note that the kink in $V_\mathrm{foam}$, which appears at $V_\mathrm{foam}\approx$\SI{15}{\milli\liter} in all curves, is a systematic instrument effect.
It is attributed to the viewing geometry of the CCD1 camera, by which the foam volume is determined.
The CCD1 camera is installed at a fixed position relative to the foam cell.
At early times, when the foam volume is still small, the camera views the foam from a different angle and partially detects the foam surface from above.
As the developing foam rises in the cell, the viewing geometry changes and the foam starts to be detected from the side.
This change in viewing geometry affects the apparent foam height and, thus, the detected foam volume.

Please note that these foams were generated with a bulk liquid volume of \SI{40}{\milli\liter} of \SI{0.075}{wt\percent} of MG5 and a frit with porosity P1.6 ((ISO 4793), pore size \SIrange{1}{1.6}{\micro\meter}).

\newpage
\section{Liquid fraction during foam formation}
\begin{figure} [htb]
\centering
  \includegraphics[height=5cm]{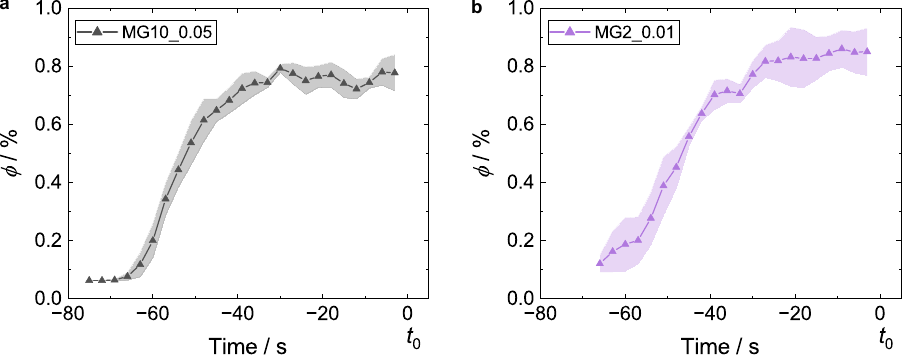}
  \caption{Time evolution of the liquid fraction $\phi$ in the foam during foam formation of samples a) MG10\_0.05 and b) MG2\_0.01. Foam formation is completed at $t_\mathrm{0} = \SI{0}{\second}$.}
  \label{fgr:LF_Formation}
\end{figure}

In Fig.~\ref{fgr:LF_Formation} the time evolution of the liquid fraction $\phi$ in the foam  of samples a) MG10\_0.05 and b) MG2\_0.01 during foam formation is shown, which exhibit intermittent drops during the increase of $\phi(t)$.  

\newpage
\section{Fit parameters of the liquid fraction during foam collapse}

\begin{figure} [htbp!]
\centering
  \includegraphics[height=15cm]{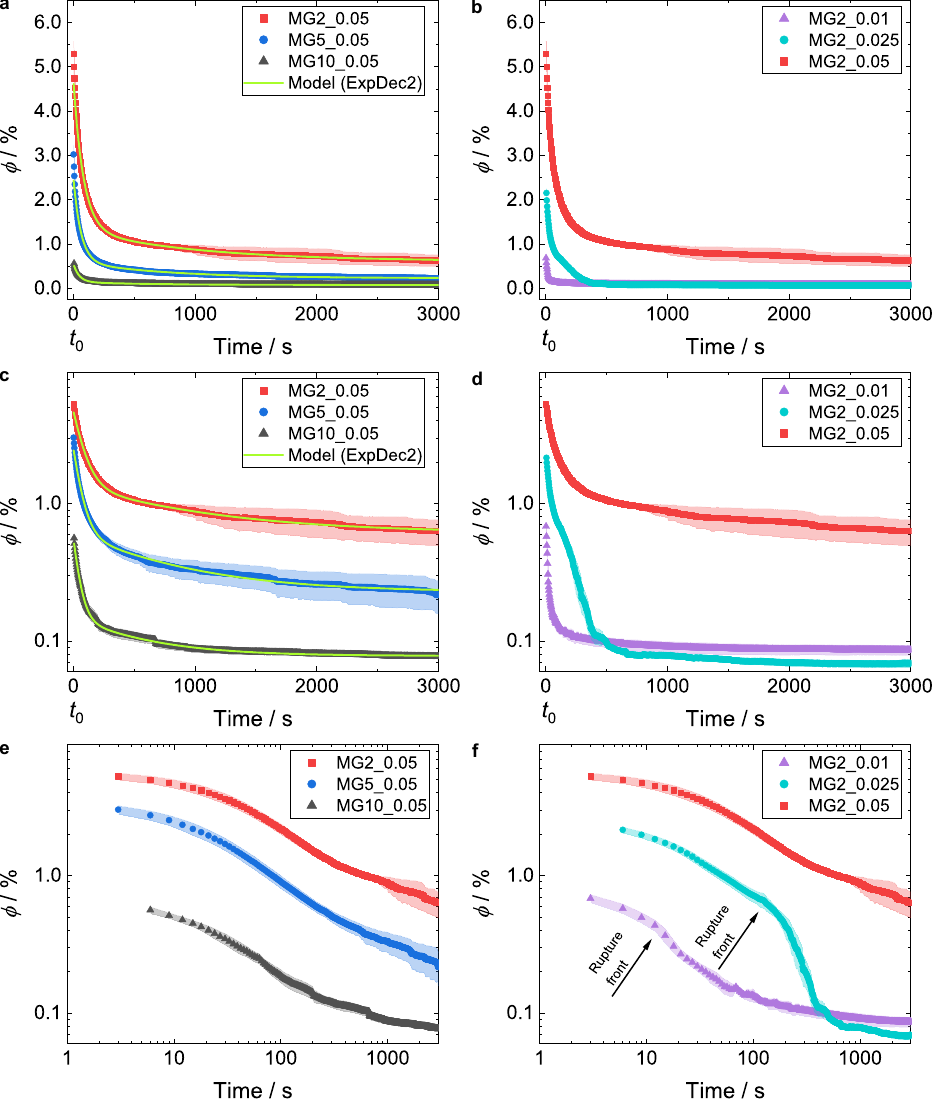}
  \caption{Time evolution of the liquid fraction $\phi$ in the foam after $t_\mathrm{0}$ as a function of a), c) and e) \mbox{cross-linker} density and b), d) and f) MG concentration, shown on \mbox{lin-lin}, \mbox{log-lin} and \mbox{log-log} scales. Symbols represent the experimental data, while the green solid lines represent fits according to a \mbox{bi-exponential} decay model. In addition the onsets of the rupture front observed for samples MG2\_0.01 and MG2\_0.025 are indicated by arrows.}
  \label{fgr:log_lin}
\end{figure}

The time evolution of the liquid fraction $\phi$ in the foam during foam collapse is investigated after $t_\mathrm{0} = \SI{0}{\second}$.
The decrease of $\phi(t)$ is shown in Fig.~\ref{fgr:log_lin} as a function of \mbox{cross-linker} density and MG concentration on a \mbox{lin-lin} scale in panels a) and b), respectively, on a \mbox{log-lin} scale in panels c) and d), respectively, and on an a \mbox{log-log} scale in panels e) and f), respectively.

The decrease in $\phi(t)$ as a function of \mbox{cross-linker} density was fitted by a \mbox{bi-exponential} decay model according to Eq.~\ref{eq:bi_exp}

\begin{equation}
    y(x) = A_1 \exp\left[-\frac{x'}{\tau_1}\right]
    + A_2 \exp\left[-\frac{x'}{\tau_2}\right]+y_0 
\label{eq:bi_exp}
\end{equation}

With $\quad x' = x - x_\mathrm{ref}$, where $x_\mathrm{ref}$ denotes the first data point included in the fit, corresponding to the first value in the data series after $t_\mathrm{0}$.

The corresponding fit parameters are listed in Tab.~\ref{tab:bi_ex_LF}.
Furthermore, the fit is indicated by a green solid line in Fig.~\ref{fgr:log_lin}a) and c).

\begin{table*}[h!]
\small
  \caption{Fit parameters of bi-exponential decay model of the time evolution of the liquid fraction $\phi$ in the foam after $t_\mathrm{0}$ as a function of \mbox{cross-linker} density}
  \label{tbl:bi_exp_para}
  \begin{tabular*}{\textwidth}{@{\extracolsep{\fill}}llllll}
    \hline
    Foam & $A_\mathrm{1}$ & $\tau_\mathrm{1}$ & $A_\mathrm{2}$ & $\tau_\mathrm{2}$  & $y_\mathrm{0}$\\
    \hline
    MG2\_0.05 & $3.272 \pm 0.018$ & $81.55 \pm 0.50$ & $0.7338 \pm 0.0029$ & $969 \pm 13$ & $0.6129 \pm 0.0027$ \\
    MG5\_0.05 & $1.829 \pm 0.017$  & $60.89 \pm 0.68$ & $0.3782\pm 0.0034$ & $757 \pm 13$ & $0.2285
 \pm 0.0012$  \\
    MG10\_0.05 & $0.3662 \pm 0.0039$ & $47.13 \pm 0.70$ & $0.06899 \pm 0.00083$ &  $557.5 \pm 6.2$ & $0.077907\pm 0.000065$ \\
    \hline
    \label{tab:bi_ex_LF}
  \end{tabular*}
\end{table*}

In addition, the \mbox{log-log} representation of $\phi(t)$ as a function of MG concentration shown in Fig.~\ref{fgr:log_lin}f) revealed the presence of a rupture front for samples MG2\_0.01 and MG2\_0.025 occurring at $\approx\SI{15}{\second}$ and $\approx\SI{120}{\second}$, respectively, indicated by an arrow.

